\documentclass[12pt]{article}
\usepackage{epsf}
\usepackage{cite,,epsfig}
\usepackage{graphicx}

\newcommand{\TeV}{~{\rm TeV}}

\newcommand{\be}{\begin{equation}}
\newcommand{\ee}{\end{equation}}
\newcommand{\bea}{\begin{eqnarray}}
\newcommand{\eea}{\end{eqnarray}}
\newcommand{\beq}{\begin{equation}}
\newcommand{\eeq}{\end{equation}}
\newcommand{\ba}{\begin{array}}
\newcommand{\ea}{\end{array}}
\newcommand{\beqa}{\begin{eqnarray}}
\newcommand{\eeqa}{\end{eqnarray}}
\newcommand{\dis}{\displaystyle}

\newcommand{\cO}{{\cal O}}

\newcommand{\no}{\nonumber}

\newcommand{\lsim}{\stackrel{<}{_\sim}}
\newcommand{\gsim}{\stackrel{>}{_\sim}}

\newcommand{\BR}{{\mathcal B}}
\newcommand{\cB}{{\mathcal B}}

\newcommand{\Btaun}{{B \to \tau \nu}}

\newcommand{\tchi}{{\tilde \chi}}
\begin{document}

\thispagestyle{empty}
\begin{flushright}
February 2007
\end{flushright}
\vskip 1.5 true cm 

\begin{center}

{\Large\bf Flavour physics at large \boldmath$\tan\beta$ with a Bino-like LSP} 
  \\ [25 pt]
{\sc {\sc G. Isidori${}^{a}$, F. Mescia${}^{a}$, P. Paradisi${}^{b}$, D. Temes${}^{a}$ } 
  \\ [25 pt]
{\sl ${}^a$INFN, Laboratori Nazionali di Frascati, Via E. Fermi 40, 
           I-00044 Frascati, Italy} \\ [5 pt] 
{\sl ${}^b$Departament de F\'{\i}sica Te\`orica  and IFIC,  \\
Universitat de Val\`encia--CSIC, E--46100 Burjassot, Spain} \\ [25 pt]  }   
{\bf Abstract} \\
\end{center}
The MSSM with large $\tan\beta$ and heavy  squarks
($M_{\tilde q} \gsim 1$~TeV) 
is a  theoretically well motivated and phenomenologically
interesting extension of the SM.  This scenario
naturally satisfies all the 
electroweak precision constraints and, in the case 
of not too heavy  slepton sector  ($M_{\tilde \ell} \lsim 0.5$~TeV), 
can also easily accommodate the $(g-2)_\mu$ anomaly.
Within this framework non-standard effects could possibly 
be detected in the near future in a few 
low-energy flavour-violating observables, such as $\BR(\Btaun)$,
$\BR(B_{s,d}\to \ell^+\ell^-)$, $\BR(B\to X_s \gamma)$, and
 $\BR(\mu\to e \gamma)$. Interpreting the $(g-2)_\mu$ anomaly 
as the first hint of this scenario, we analyse the correlations
of these low-energy observables 
under the additional assumption
that the relic density of a Bino-like LSP 
accommodates the observed dark matter distribution. 
\noindent

\vskip 1.5 cm

\section{Introduction}
Within the Minimal Supersymmetric extension of the Standard Model (MSSM),
the scenario with large $\tan\beta$ and heavy squarks is
a particularly interesting subset of the parameter space.
On the one hand, values of $\tan\beta \sim$ 30--50
can allow the unification of top and bottom Yukawa couplings,
as predicted in well-motivated grand-unified models \cite{GUT}.
On the other hand, heavy soft-breaking terms in the squark sector
(both bilinear and trilinear couplings) with large $\tan\beta$
and a Minimal Flavour Violating (MFV) structure \cite{MFV0,MFV} lead
to interesting phenomenological virtues. On the one hand, this 
scenario can easily accommodate all the existing constraints from 
electroweak precision tests and flavour physics. In particular, in 
a wide region of the parameter space, the lighetst Higgs boson mass is above the present 
exclusion bound. On the other hand,  
if the slepton sector is not too heavy, within this framework one can 
also find a natural description of the
present $(g-2)_\mu$ anomaly. In the near future,  
additional low-energy signatures of this scenario 
could possibly show up in $\BR(\Btaun)$,
$\BR(B_{s,d}\to \ell^+\ell^-)$, and $\BR(B\to X_s \gamma)$
(see Ref.~\cite{IP,vives} for a recent phenomenological discussion).
In the parameter region relevant to $B$-physics 
and the $(g-2)_\mu$ anomaly, also a few Lepton Flavor 
Violating (LFV) processes 
(especially $\mu \to e \gamma$) are generally predicted 
to be within the range of upcoming experiments.  
In this paper we analyse the correlations of the most interesting low-energy 
observables of this scenario 
under the additional assumption that the relic density of a Bino-like lightest
supersymmetric particle (LSP) accommodates the observed dark matter 
distribution  (the constraints and reference ranges for the low-energy 
observables considered in this work can be 
found in Sect.~\ref{sect:combined}).

Recent astrophysical observations consolidate the hypothesis
that the universe is full of dark matter localized in 
large clusters \cite{wmap}. The cosmological density of 
this type of matter is determined with good accuracy 
\begin{equation}
0.079 \leq \Omega_{\rm CDM} h^2 \leq 0.119 \quad \rm{at}\: 2\sigma\:
\rm{C.L.},  
\label{eq:omg}
\end{equation}
suggesting that it
is composed by stable and  weakly-interactive 
massive particles (WIMPs).
As widely discussed in the literature (see e.g.~Ref.~\cite{Profumo}
for recent reviews), in the MSSM with $R$-parity conservation a perfect candidate 
for such form of matter is the neutralino (when it turns out 
to be the LSP) \cite{neutralino_LSP}. 
In this scenario, due to the large amount of LSP produced in the early universe, 
the lightest neutralino must have a sufficiently large annihilation cross-section
in order to satisfy the upper bound on the relic abundance.

\begin{figure}[t]
\begin{center}
\includegraphics[scale=1.0]{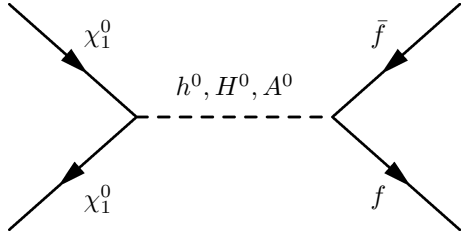}
\caption{\label{fig:1} Higgs-mediated 
neutralino annihilation amplitude.} 
\end{center}
\end{figure}

If the $\mu$ term 
is sufficiently large (i.e.~in the regime where the interesting Higgs-mediated 
effects in flavour physics are not suppressed) and $M_1$ is the lightest gaugino mass
(as expected in a GUT framework), the lightest neutralino is mostly a Bino. 
Due to the smallness of its couplings, a Bino-like LSP tends to have a 
very low annihilation cross section.\footnote{~If the conditions on $\mu$ and $M_1$ are 
relaxed, the LSP can have a dominant Wino or Higgsino component and a naturally 
larger annihilation cross-section. This scenario, which is less 
interesting for flavour physics, will not be analysed in this work.}
However, as we will discuss in Section~\ref{sect:relic}, in the regime with  
large $\tan\beta$ and heavy squarks the  relic-density 
constraints can easily be satisfied. 
In particular, the largest region of the parameter space yielding 
the correct LSP abundance is the so-called 
$A$-funnel region \cite{funnel}. Here the dominant neutralino annihilation 
amplitude is the Higgs-mediated diagram in Fig.~\ref{fig:1}. 
Interestingly enough, in this case several of the parameters
which control the amount of relic abundance, such as
$\tan\beta$ and the heavy Higgs masses, also play a key role
in flavour observables. As a result, in this scenario
imposing the dark-matter constraints leads to a well-defined
pattern of constraints and correlations on the low-energy observables 
which could possibly be tested in the near future.
The main purpose of this article is the investigation of this 
scenario.

The interplay of $(g-2)_\mu$, $\BR(B_{s,d}\to \ell^+\ell^-)$, $\BR(B\to X_s \gamma)$, 
and dark-matter constraints in the MSSM have been addressed 
in a series of recent works, focusing both on 
relic abundance \cite{New_relic} and on
direct WIMPs searches \cite{New_wimps}.
Our analysis is complementary to those studies
for two main reasons: i) the inclusion of 
$\BR(\Btaun)$, which starts to play a significant 
role in the large $\tan\beta$ regime, and will become even
more significant in the near future; ii) the study of 
a phenomenologically interesting region of the  
MSSM parameter space which goes beyond the 
scenarios analysed in most previous 
studies (see Section~\ref{sect:relic}).

The plan of the paper is the following: 
in Section~\ref{sect:relic} we recall the 
ingredients to evaluate the relic density in the MSSM, and determine 
the key parameters of the interesting $A$ funnel region.
In Section~\ref{sect:low} we present a brief updated on the 
low-energy constraints on this scenario; we analyse constraints and  
correlations on the various low-energy observables after imposing the 
dark-matter constraints; we finally study the possible correlations 
between $(g-2)_\mu$ and the lepton-flavour violating decays 
$\BR(\mu\to e \gamma)$ and  $\BR(\tau\to \mu \gamma)$.
The results are summarized in the Conclusions.

\section{Relic Density}
\label{sect:relic}

In the following we assume that  relic neutralinos 
represent a sizable fraction of the observed dark matter. 
In order to check if a specific choice of the MSSM parameters is 
consistent with this assumption, we need to ensure two 
main conditions: i) the LSP is a thermally produced neutralino; 
ii) its relic density is consistent with the astrophysical observation 
reported in  Eq.~(\ref{eq:omg}). 

In the MSSM there are four neutralino mass eigenstates,
resulting from the admixture of the two neutral
gauginos ($\tilde W^0, \tilde B$) and the two neutral higgsinos 
($\tilde H_1^0, \tilde H_2^0$). The lightest neutralino can be
defined by its composition,
\begin{equation}
\tchi_1 = Z_{11} \tilde B + Z_{12} \tilde W^0 + Z_{13} \tilde H_1^0 +
Z_{14} \tilde H_2^0
\end{equation}
where the coefficients $Z_{1i}$ and the mass eigenvalue ($M_{\tchi_1}$) 
are determined by the diagonalization of the mass matrix
\begin{equation}
\label{neumass}
\mathcal{M}_{\tchi} =
\left( \begin{array}{cccc}
M_1             &0            &-m_Z \cos\beta\, s_W & m_Z \sin\beta\, s_W \\
0               &M_2          & m_Z \cos\beta\, c_W &-m_Z \sin\beta\, c_W \\
-m_Z \cos\beta\, s_W  & m_Z \cos\beta\, c_W &0            &-\mu           \\
 m_Z \sin\beta\, s_W  &-m_Z \sin\beta\, c_W &-\mu         &0     \end{array}
\right).
\end{equation}
As usual, $\theta_W$ denotes the weak mixing angle ($c_W \equiv\cos\theta_W$,
$s_W \equiv\sin\theta_W$) and $\beta$ is defined by the relation 
$\tan\beta \equiv v_2 / v_1$, where $v_{2(1)}$ is the vacuum expectation 
value of the  Higgs coupled to up(down)-type quarks; $M_1$ and $M_2$ are 
the soft-breaking gaugino masses and $\mu$ is the supersymmetric-invariant 
mass term of the Higgs potential.

In order to compute the present amount of neutralinos  
we assume a standard thermal history of the universe \cite{history} 
and evaluate the annihilation and coannihilation cross-sections using the 
micrOMEGAs~\cite{micromegas} code. Since we cannot 
exclude other relic contributions in addition to the neutralinos, 
we have analysed only the consistency with the upper limit 
in Eq.~(\ref{eq:omg}). This can
be translated into a lower bound on the neutralino cross sections: the
annihilation and coannihilation processes have to be effective 
enough to yield a sufficiently low neutralino density at present time.

With respect to most of the existing analysis of dark-matter 
constraints in the MSSM, in this work we do not impose relations 
among the MSSM free parameters dictated by specific supersymmetry-breaking  
mechanisms. Consistently with the analysis of Ref.~\cite{IP}, we
follow a bottom-up approach supplemented by few underlying hypothesis,
such as the large value of $\tan\beta$ and the heavy soft-breaking terms 
in the squark sector. As far as the neutralino mass terms are concerned, 
we employ the following two additional hypotheses: the GUT 
relation  $M_1 \approx M_2/2 \approx M_3/6$, and the relation $\mu > M_1$, which 
selects the parameter region with the most interesting 
Higgs-mediated effects in flavour physics 
(see Section~\ref{sect:low}).\footnote{~These two assumptions are not strictly 
necessary.  From this point of view, 
our analysis should not be regarded as the most general analysis 
of dark-matter constraints in the MSSM at large $\tan\beta$.
We employ these assumptions both to reduce the number of 
free parameters and to maximize the potentially visible 
non-standard effects in the flavour sector. In particular, 
the condition  $\mu > M_1$ does not follow from model-building 
considerations (although well-motivated scenarios, such as
mSUGRA, naturally predict $\mu > M_1$ in large portions 
of the parameter space), rather from the requirement of non-vanishing 
large-$\tan\beta$ effects in $B\to\mu^+\mu^-$ and other 
low-energy observables \cite{Bmmth,Babu,Buras}
(which provide a distinctive signature of this scenario). }
These two hypotheses imply that the lightest 
neutralino is Bino-like (i.e.~$Z_{11} \gg Z_{1j\not=1}$)
with a possible large Higgsino fraction when $\mu=\mathcal{O}(M_1)$.
Due to the smallness of the $\tilde B$ couplings, some
enhancements of the annihilation and coannihilation processes 
are necessary in order to fulfill the relic density constraint. 
In general, these enhancements can be produced by the following 
three mechanisms \cite{darkl,Profumo}:
\begin{itemize}
\item
Light sfermions. For light sfermions,
the $t$-channel sfermion exchange leads
to a sufficiently large annihilation amplitude
into fermions with large hypercharge.
\item
Coannihilation with other SUSY particles. If the next-to-lightest
supersymmetric particle (NLSP) mass is closed to $M_{\tchi_1}$, the 
coannihilation process NLSP+LSP~$\to$~SM can be efficient enough to
reduce the amount of neutralinos down to the allowed range.
A relevant coannihilation process in our
scenario occurs when the NLSP is the lightest stau lepton 
(stau annihilation region). This mechanism becomes relevant when 
the lightest stau mass, 
$M^{2}_{\tilde{\tau}_{R}}\approx M^{2}_{\tilde \ell}-m_{\tau}\mu\tan\beta$, satisfies 
the following condition
\beq
M_{\tchi_1} < M_{\tilde{\tau}_{R}}\lsim 1.1 \times  M_{\tchi_1 }~.
\label{eq:staucond}
\eeq
Other relevent coannihilation processes take place 
when $\mu$ is sufficiently close to $M_1$. 
In this case the LSP 
coannihilation with a light neutralino or chargino (mostly higgsino-like
and thus with mass $M_{\tchi_2^0,\tchi_1^+} \sim \mu$), can become 
efficient.

\item
Resonant processes. Neutralinos can efficiently annihilate into down-type
fermion pairs through $s$-channel exchange close to resonance (see Fig.~\ref{fig:1}).
At large $\tan\beta$, the potentially dominant effect is through the heavy-Higgs
exchange ($A$ and $H^0$) and in this case the resonant condition 
implies 
\beq
M_{\tchi_1}\approx M_A/2~.
\eeq
At resonance the amplitude is proportional to 
$(M_{\tchi_1}/M^2_A)\,(Z_{11} Z_{13,14}) (m_{b,\tau}/m_W) \tan\beta$
which shows that the lightest neutralino must have a non-negligible 
higgsino component ($Z_{13,14} \not=0$), and that the annihilation 
into  $b$ and $\tau$ fermions grows at large $\tan\beta$ 
(relaxing the resonance condition).
\end{itemize}

Because of the  heavy squark masses,
the first of these mechanisms is essentially excluded 
in the scenario we are considering: we assume squark 
masses in the $1-2$ TeV range and, in order to maintain a natural ratio 
between squark and slepton masses, this implies
sleptons masses in the $0.3$--$1$ TeV range.
The second mechanism can occur, but only in specific regions.
 On the other hand,
the $s$-channel annihilation
$\tchi \tchi \to H, A \to b\bar b (\tau^+ \tau^-)$ can be very
efficient in a wide region of the parameter space of
our scenario.

\begin{figure}[t]
\begin{center}
\includegraphics[scale=0.65]{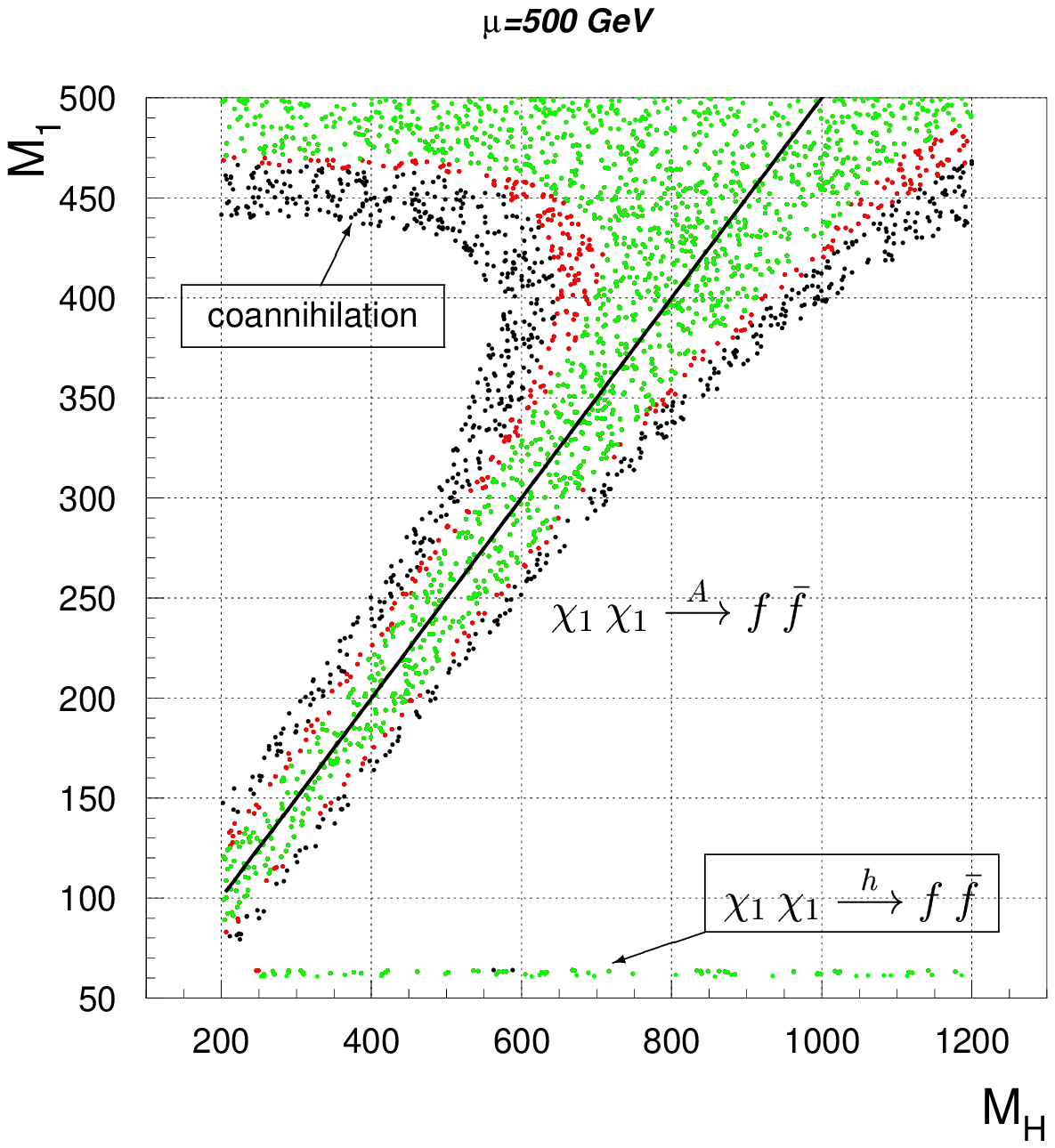}
\includegraphics[scale=0.65]{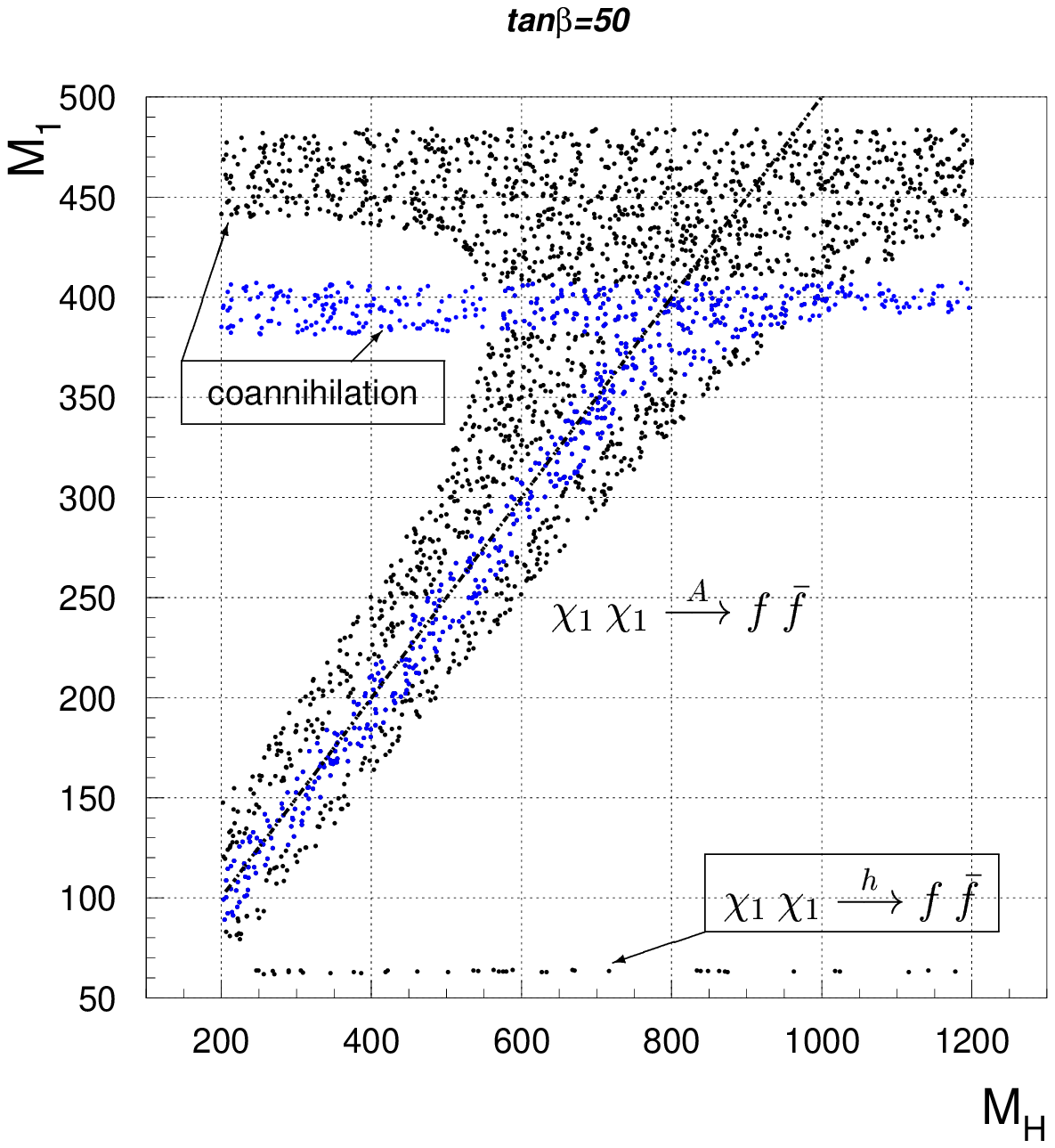} \\
\caption{\label{fig:mu500}
Allowed regions in the $M_1$--$M_H$ plane
satisfying the relic density constraint $\Omega h^2 < 0.119$
for $M_{\tilde q} = 2 M_{\tilde \ell} = |A_U| =1$~ TeV.
Left panel: $\mu=0.5\TeV$ with $\tan\beta=20$ (green),
$30$ (green+red) and $50$ (all points
up to $M_1 \approx 480$ GeV, see right panel).
Right panel: $\mu=0.5 \TeV$ (black) and  $\mu=1 \TeV$ (blue)
for $\tan\beta=50$ (a color version of this and the
following figures can be found in the on-line version of this
article).}
\end{center}
\end{figure}

In Fig.~\ref{fig:mu500} we explore the dark-matter constraints 
in the $M_1$--$M_H$ plane, assuming heavy squarks and sleptons
($M_{\tilde q} = 1$ TeV, $M_{\tilde \ell} = 0.5$ TeV) and
large trilinear couplings ($|A_U|= 1$ TeV). The allowed points have been
obtained for different values of $\mu$ and $\tan\beta$. The dependence
on $\tan\beta$ at fixed $\mu$ ($\mu =0.5$ TeV) is illustrated by the left panel,
while the $\mu$ dependence at fixed $\tan\beta$ ($\tan\beta=50$)
is illustrated by the right panel. In all cases the heavy-Higgs 
resonant region, $M_1 \approx M_{H,A}/2$, is the most important 
one.\footnote{~We recall that for sufficiently heavy $M_H\geq 300 \rm{GeV}$, 
the heavy Higgses are almost degenerate: $M_H \approx M_A$.   
We also recall that, within mSUGRA models, $M_H$, $M_1$
and $\tan\beta$ are not independent parameters. 
In this case, the $A$-funnel condition $M_H \approx 2 M_1$ 
is achieved only in the very large $\tan\beta$ regime $45 < \tan\beta < 60$.
In our scenario, where $M_H$ and $M_1$ are assumed to 
be free parameters, this constraints is relaxed and smaller values
of $\tan\beta$ are also allowed.}
The $M_H$-independent regions for $M_1 > 450$ GeV and $M_1 \sim 60$ GeV
are generated by the  $\tchi$ coannihilation mechanisms and the $h$ 
resonance amplitude, respectively. As can be seen, in the heavy-Higgs resonant case
the allowed region becomes larger for larger $M_H$ values: this is because
the Higgs width grows with $M_H$ and therefore the resonance region becomes larger.
For a similar reason, and also because the annihilation cross sections
grow with $\tan\beta$, the  allowed region becomes larger for larger $\tan\beta$ values.
As far as the $\mu$ dependence is concerned, the heavy-Higgs resonance region is larger
for small $\mu$ values. This is because the $\tchi \tchi A$ coupling, relevant in the
resonant process, depends on the Higgsino component of $\tchi$: for large $\mu$,
$\tchi$ is almost a pure $\tilde B$ and the $\tchi \tchi A$ 
coupling is suppressed.
This fact can be used to set a theoretical upper 
limit on the $\mu$ parameter in this specific 
framework: $\mu$ must be larger than $M_1$ in order to reproduce a Bino LSP,
but it should not be too heavy not to suppress too much the Bino annihilation amplitude.

Notice that in the right panel of Fig.~\ref{fig:mu500}
only the $\tchi-\tilde \tau$ coannihilation process
is active when $\mu=1~\rm{TeV}$. On general grounds, given a
left-handed slepton mass $M_{\tilde{\ell}}$, the stau coannihilation region 
appears for lower $M_1$ if $\mu$ increases, since $M_{\tilde{\tau}}$ decreases 
with increasing $\mu$.
Notice also that the $h$ resonance region disappears for large $\mu$, due
to the smallness of the $\tchi \tchi h$ coupling. In both figures points with
$M_{\tilde{\tau}} <  M_{\tchi_1}$ have not been plotted since they are ruled out.

In summary, the MSSM scenario we are considering is mainly motivated by 
flavour-physics and electroweak precision observables. As we have shown in this 
Section, in this framework the dark matter constraints can be easily fulfilled 
with a Bino-like LSP and an efficient Higgs-mediated Bino annihilation amplitude.
The latter condition implies a strong link between the gaugino and 
the Higgs sectors (most notably via the relation $M_1 \approx M_H/2$).
This link reduces the number of free parameters, enhancing the 
possible correlations among low-energy observables.

\section{Low-energy observables}
\label{sect:low}

In this Section we analyse the correlations of new-physics effects in  
$a_\mu=(g-2)_\mu/2$,  $\BR(\Btaun)$, $\BR(B_{s,d}\to \ell^+\ell^-)$, 
$\BR(B\to X_s \gamma)$, $\BR(\mu\to e \gamma)$, and  $\BR(\tau\to \mu \gamma)$, 
after imposing the dark matter constraints. As far as the $B$-physics observables 
are concerned, we use the existing calculations of supersymmetric 
effects in the large $\tan\beta$ regime
which have been been recently 
reviewed in Ref.~\cite{IP,vives}.\footnote{~See in 
particular Ref.~\cite{bsgamma}
for $B\to X_s\gamma$, Ref.~\cite{Bmmth,Babu,Buras} for $\BR(B_{s,d}\to \ell^+\ell^-)$,
Ref.~\cite{IP,Hou} for $\BR(\Btaun)$, and Ref.~\cite{gm2th} for $(g-2)_\mu/2$. 
After this work was completed, 
a new theoretical analysis of  large $\tan\beta$ effects in $B$ physics, within the MFV-MSSM, 
has appeared \cite{Haisch}. As shown in  Ref.~\cite{Haisch}, the 
renormalization of both $\tan\beta$ and the Higgs masses 
may lead to sizable modifications of the commonly adopted 
formulae for $\Delta M_{B_{s,d}}$ (see Ref.~\cite{Buras}),
which are valid only in the 
 $M_H \gg m_W$ limit \cite{MFV}. On the numerical side, 
these new effects turn out to be non-negligible only  
in a narrow region of light $M_H$ 
($M_{A}\lsim 160$~GeV or $M_H \lsim 180$~GeV)  
which is not allowed within our analysis. These new effects are therefore 
safely negligible for our purposes.}
However, since a few inputs have changed since then, 
most notably the $\BR(\Btaun)$ measurements \cite{Btaunu_Babar,Btaunu_Belle}  
and the SM calculation of $\BR(B\to X_s \gamma)$ \cite{bsgth},
in the following we first present a brief updated on these 
two inputs. We then proceed analysing the implications 
on the MSSM parameter space of  $a_\mu$ and 
$B$-physics observables after imposing the 
dark matter constraints. Finally, the possible correlations 
between $a_\mu$ and the lepton-flavour violating decays 
$\BR(\mu\to e \gamma)$ and  $\BR(\tau\to \mu \gamma)$ in 
this framework are discussed.

\subsection{Updated constraints from $B\to\tau\nu$ and $B\to X_s \gamma$}

Due to its enhanced sensitivity to tree-level charged-Higgs exchange,
\cite{Hou} $B\to\tau\nu$ is one of the most clean probes of the large
$\tan\beta$ scenario. The recent $B$-factory results~\cite{Btaunu_Babar,Btaunu_Belle}~,
\beqa
{\cal B}(\Btaun)^{\rm Babar} &=&
(0.88^{+0.68}_{-0.67}(\mbox{stat}) \pm 0.11(\mbox{syst}))\times 10^{-4}~, \no \\
{\cal B}(\Btaun)^{\rm Belle} &=& 
(1.79^{+0.56}_{-0.49}(\mbox{stat})^{+0.46}_{-0.51}(\mbox{syst}))\times 10^{-4}~,
\label{eq:btnexp}
\eeqa
leads to the average ${\cal B}(\Btaun)^{\rm exp} = (1.31 \pm 0.49)\times 10^{-4}~$.
This should be compared with the SM expectation
${\cal B}(\Btaun)^{\rm SM} = G_{F}^{2} m_{B} m_{\tau}^{2} f_{B}^{2} |V_{ub}|^{2}
(1-m_{\tau}^{2}/m_{B}^{2})^{2} /(8\pi \Gamma_B)$, whose numerical value
suffers from sizable parametrical uncertainties induced by $f_B$ and $V_{ub}$.
According to the global fit\footnote{~In Ref.~\cite{UTfit}
the value of $f_B$ is indirectly determined taking into account the
information from both $B_d$--$\bar B_d$ and $B_s$--$\bar B_s$ mixing.}
of Ref.~\cite{UTfit},
the best estimate is  ${\cal B}(\Btaun)^{\rm SM} = (1.41 \pm 0.33)\times 10^{-4}~$,
which implies
\be
R_{B\tau\nu}^{\rm exp} =
\frac{\BR^{\rm exp}(\Btaun)}{\BR^{\rm SM}(\Btaun)}
~=~ 0.93 \pm 0.41~.
\label{eq:Rtn_exp}
\ee
A similar (more transparent) strategy to 
minimize the error on ${\cal B}(\Btaun)^{\rm SM}$ is the direct 
normalization of ${\cal B}(\Btaun)$ to $\Delta M_{B_d}$, given that 
$B_d$--$\bar B_d$ is not affected by new physics in our scenario~\cite{IP}.
In this case, using $ B_{B_d}(m_b) = 0.836 \pm 0.068$ 
and $|V_{ub}/V_{td}|=0.473 \pm 0.024$  \cite{UTfit}, we get
\bea
\left(R^\prime_{B\tau\nu}\right)^{\rm exp} &=& 
\frac{\BR^{\rm exp}(\Btaun)/\Delta M_{B_d}^{\rm exp}}{\BR^{\rm SM}(\Btaun)/\Delta M_{B_d}^{\rm SM}}
\\
&=& 1.27 \pm 0.50 ~=~ 1.27 \pm 0.48_{\rm exp} \pm 0.10_{|B_{B_d}|} \pm 0.13_{|V_{ub}/V_{td}|}~,
\label{eq:Rtn_prime}
\eea
in reasonable agreement with Eq.~(\ref{eq:Rtn_exp}).
Although perfectly compatible with 1 (or with no new physics contributions), 
these results leave open the possibility of $O(10\%-30\%)$ negative corrections 
induced by the charged-Higgs exchange. The present error on 
$R^{(\prime)}_{B\tau\nu}$ is too large to  provide a significant 
constraint in the MSSM parameter space. In order to illustrate the possible role
of a more precise determination of $\BR^{\rm exp}(\Btaun)$, in the following 
we will consider the impact of the reference
range $0.8 < R_{B\tau\nu} < 0.9$.
In the next 2-3 years, at the end of the $B$-factory programs, we can 
expect a reduction of the experimental error on ${\cal B}(\Btaun)$ of a factor
of 2-3. Depending on the possible shift of the central value 
of the measurement [note the large spread among the two central values in 
Eq.~(\ref{eq:btnexp})] the upper bound $R_{B\tau\nu} < 0.9$ could become 
the true 68\% or 90\% CL limit.

\medskip 

The $B\to X_s \gamma$ transition is particularly sensitive to
new physics. However, contrary to $\Btaun$, it does not receive
tree-level contributions from the Higgs sector.
The one-loop charged-Higgs amplitude, which increases the
rate compared to the SM expectation, can be partially compensated 
by the chargino-squark amplitude even for squark masses of $O(1~{\rm TeV})$.
According to the recent NNLO analysis of Ref.~\cite{bsgth},
the SM prediction is
\be
{\cal B}(B\to X_s \gamma; E_\gamma > 1.6~{\rm GeV})^{\rm SM}
= (3.15 \pm 0.23) \times 10^{-4}~,
\ee
to be compared with the experimental average~\cite{hfag,belle,babar}
\be
{\cal B}(B\to X_s \gamma;   E_\gamma > 1.6~{\rm GeV}))^{\rm exp}
 = (3.55 \pm 0.24) \times 10^{-4}~.
\ee
Combining these results, we obtain the following 1$\sigma$ CL interval
\be
1.01 < R_{ Bs\gamma} =
 \frac{ {\cal B}^{\rm exp}(B\to X_s \gamma)}{{\cal B}^{\rm SM}(B\to X_s \gamma)} < 1.24
\label{eq:Bsgamma}
\ee
which will be used to constrain the MSSM parameter space 
in the following numerical analysis.\footnote{~A slightly larger (and less 
standard) range is obtained taking into account the corrections 
associated to the $E_\gamma$ cut 
in Ref.~\cite{neubert}. For simplicity, in our numerical analysis 
we have used Eq.~(\ref{eq:Bsgamma}) as reference range. The 
 $B\to X_s \gamma$ rate in the MSSM has been evaluated using the approximate 
numerical formula of Ref.~\cite{lunghibsg}, which partially takes into account 
NNLO effects.}

\subsection{Combined constraints in the MSSM parameter space}
\label{sect:combined}

\begin{figure}[p]
\includegraphics[scale=0.38]{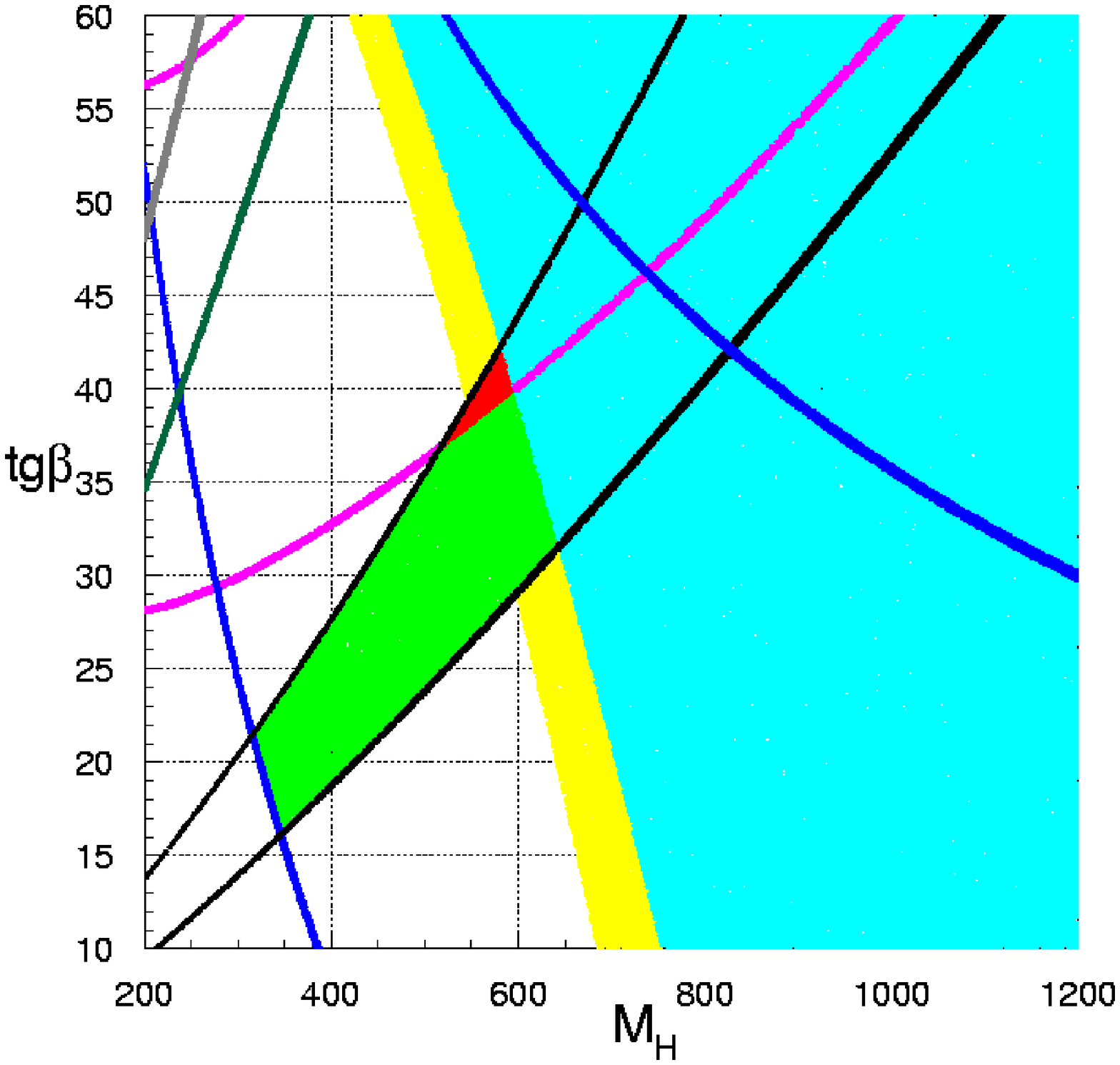}
\includegraphics[scale=0.38]{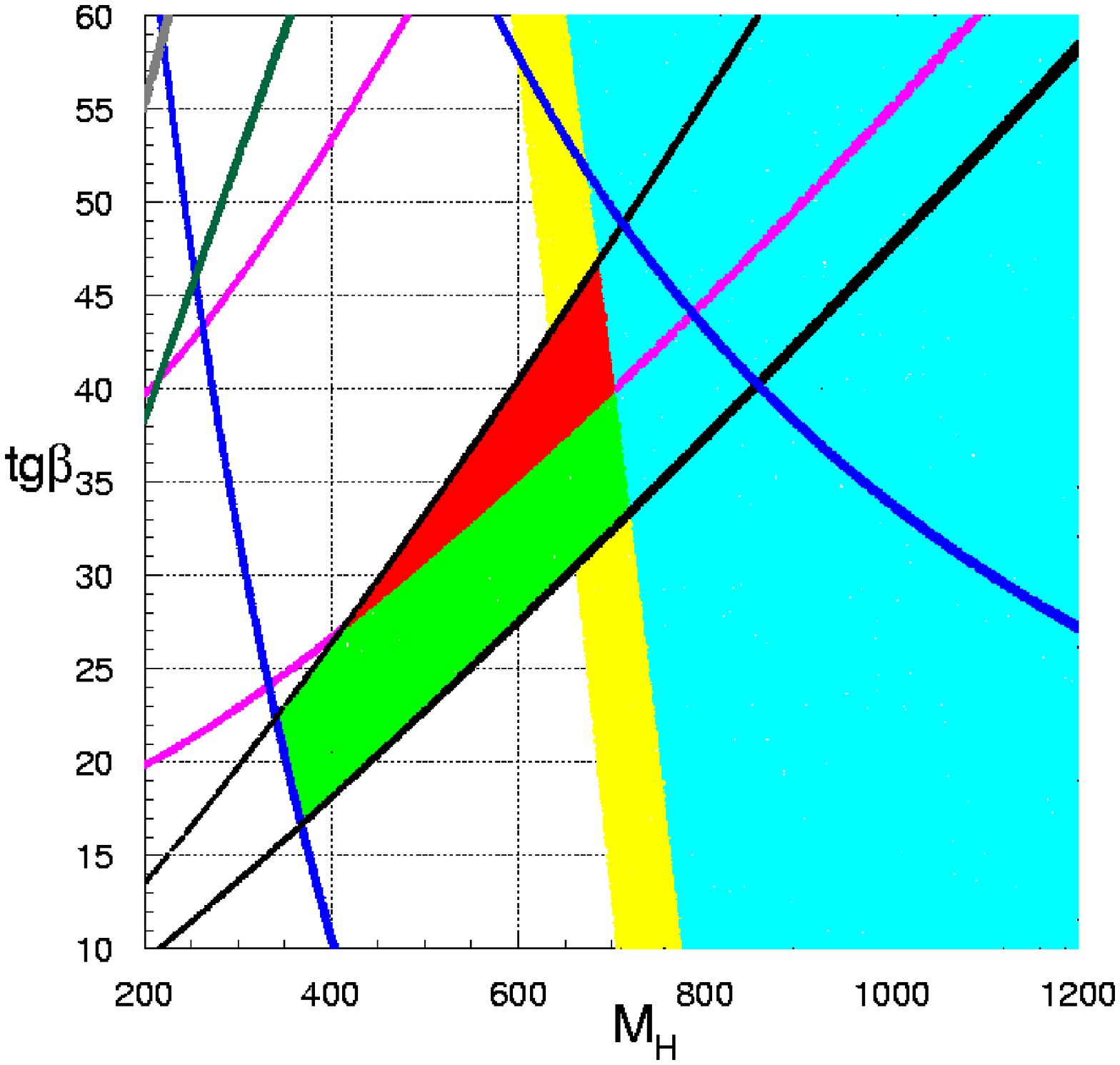} \\
\includegraphics[scale=0.38]{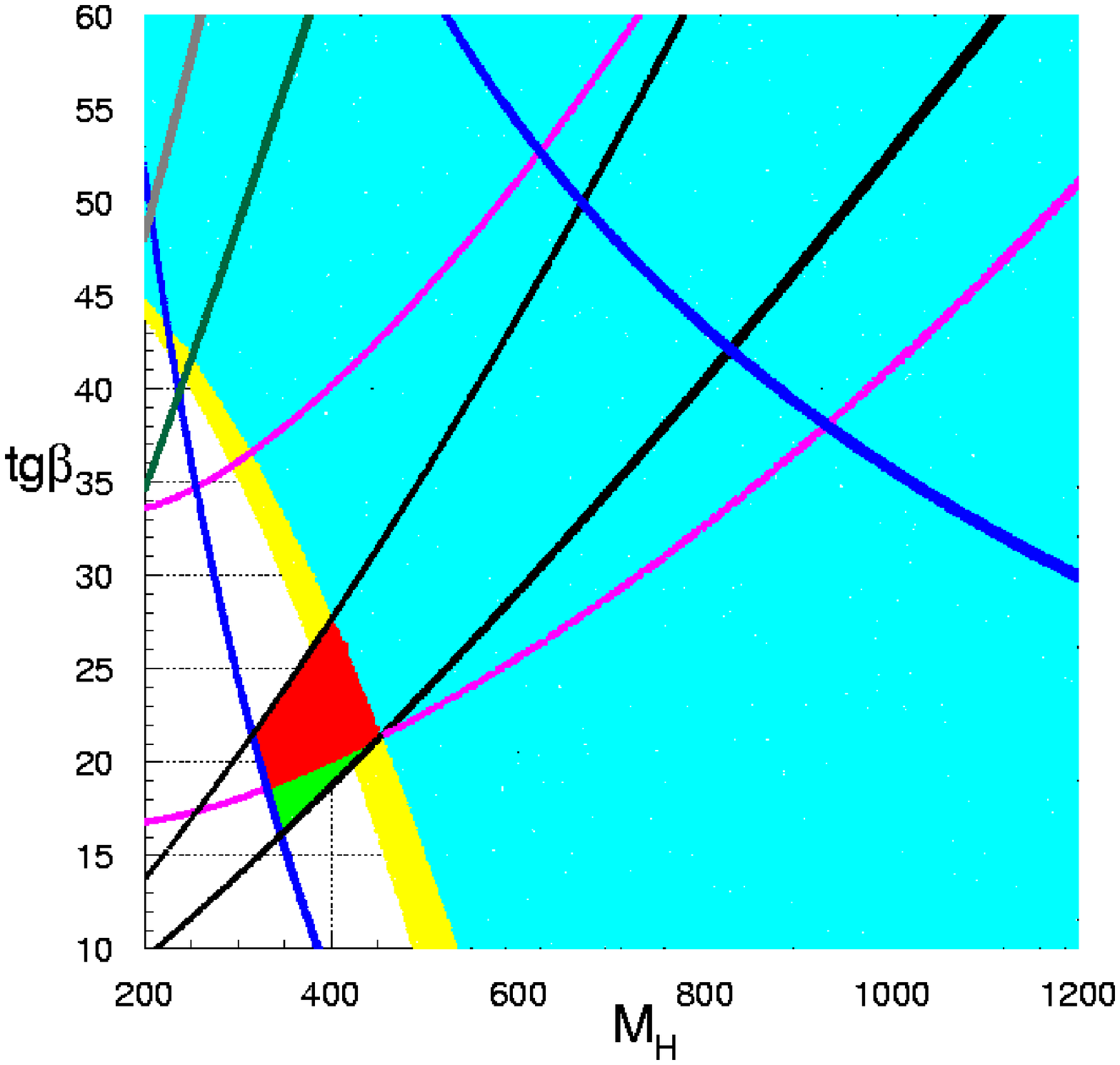}
\includegraphics[scale=0.38]{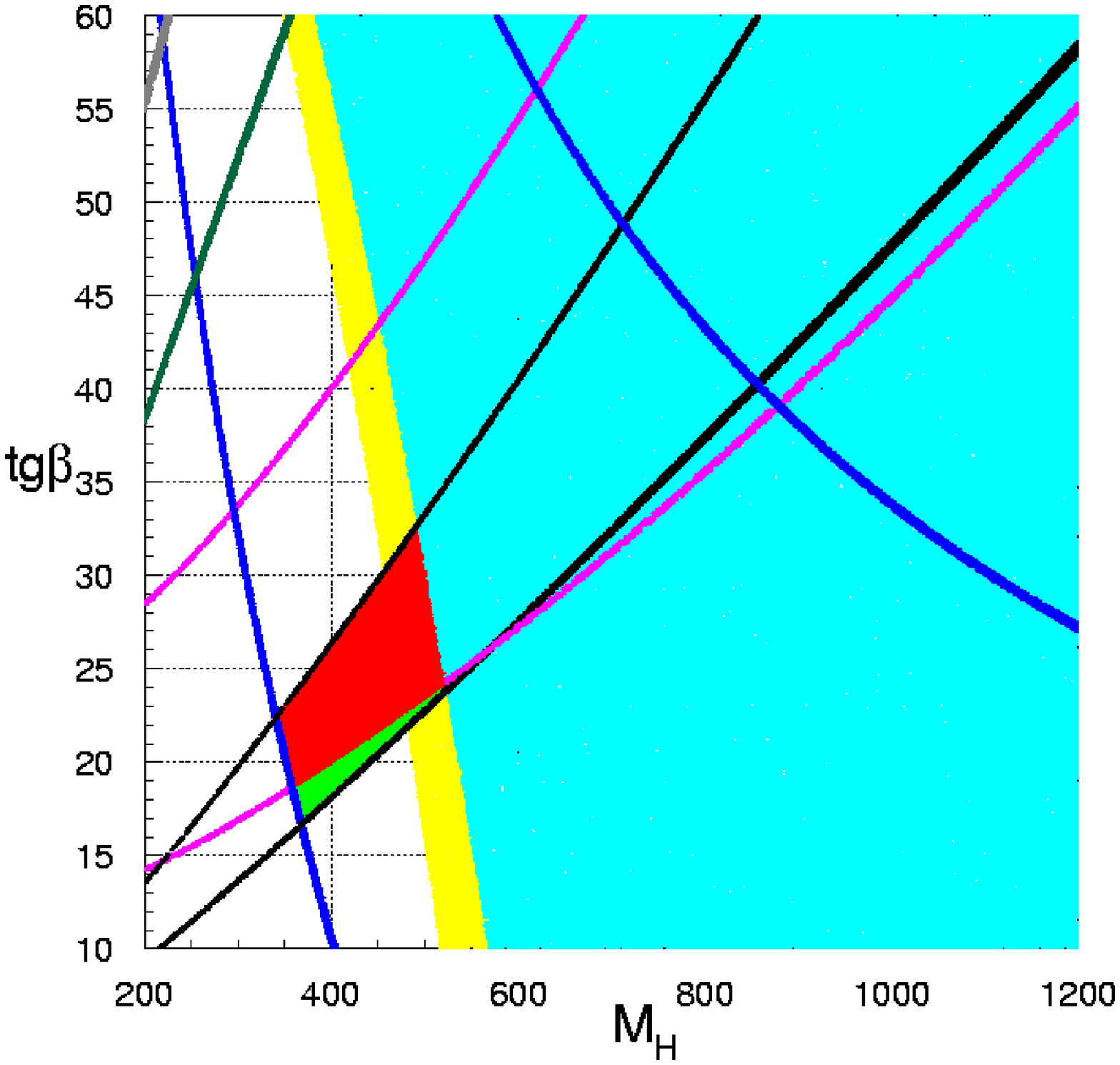}
\caption{\label{fig_MSSM_full}
Combined constraints from low-energy observables and dark matter 
in the $\tan\beta$--$M_H$ plane. The plots have been obtained for 
$M_{\tilde q} =1.5$~TeV $A_U = -1$~TeV, and $[\mu,M_{\tilde{\ell}}]=
[1.0,0.4]$~TeV (upper left); $[\mu,M_{\tilde{\ell}}]=[0.5,0.4]$~TeV (upper right);
$[\mu,M_{\tilde{\ell}}]=[1.0,0.3]$~TeV (lower left);
$[\mu,M_{\tilde{\ell}}]=[0.5,0.3]$~TeV (lower right).
The light-blue area is excluded by the dark-matter conditions.
Within the red (green) area all the reference values of the low-energy 
observables (but for $a_\mu$) are satisfied. See main text for more details.
The yellow band denote the area where the stau coannihilation mechanism 
is active ($1< M_{\tilde{\tau}_R}/M_{\tilde{B}}<1.1$); in this area
the $A$-funnel region and the stau coannihilation region overlap.}
\end{figure}

\begin{figure}[p]
\includegraphics[scale=0.38]{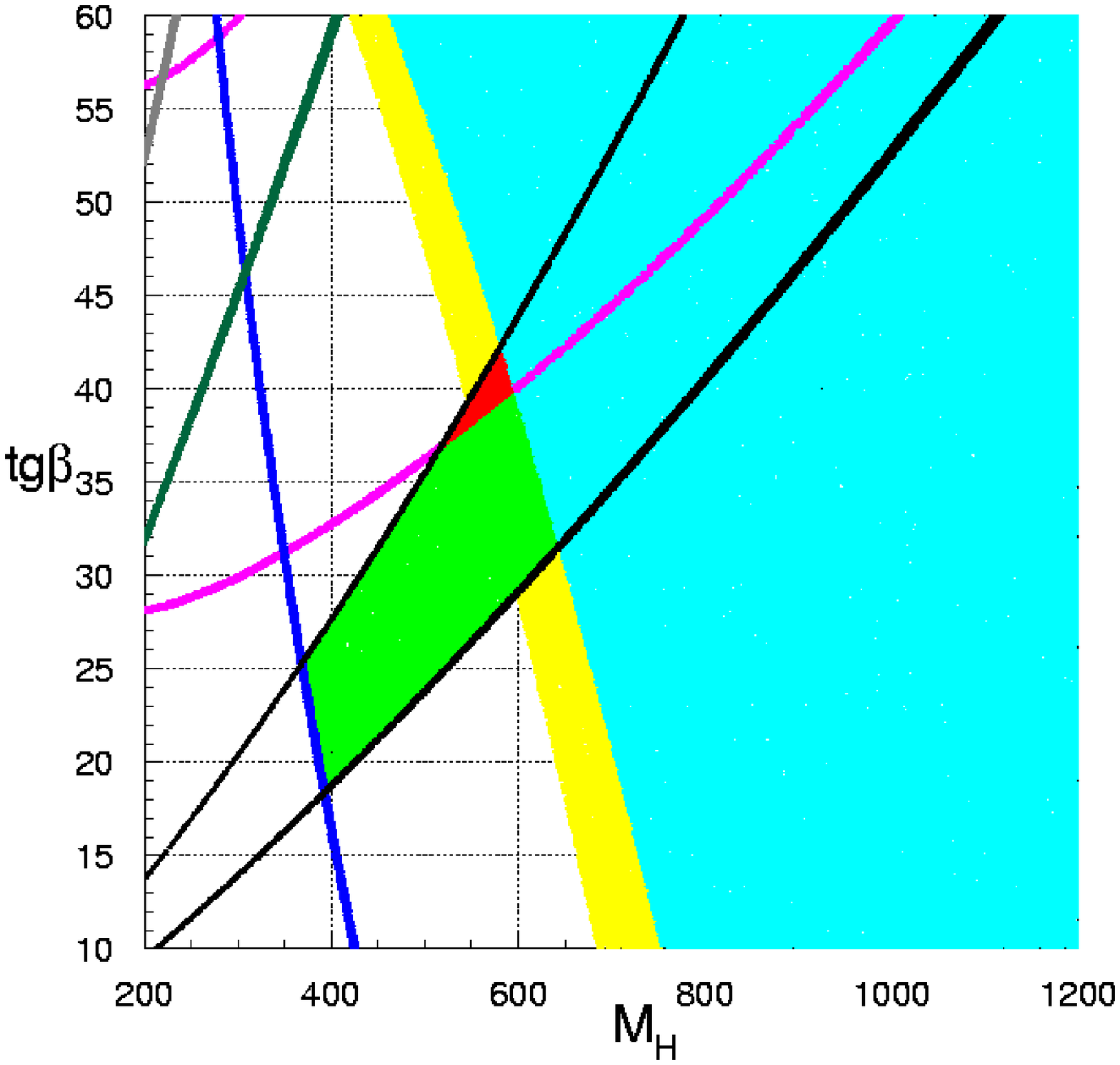} 
\includegraphics[scale=0.38]{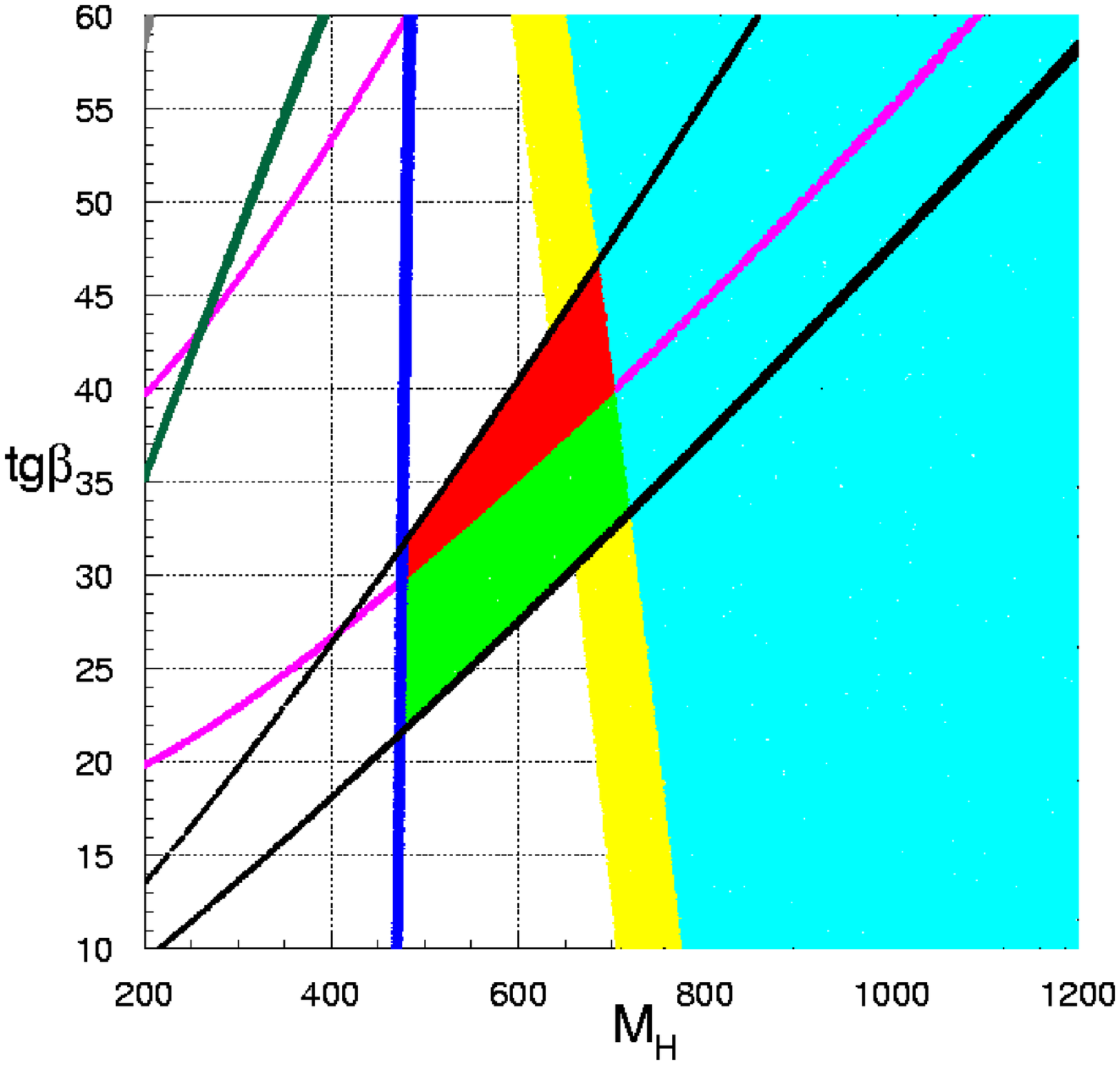}\\
\includegraphics[scale=0.38]{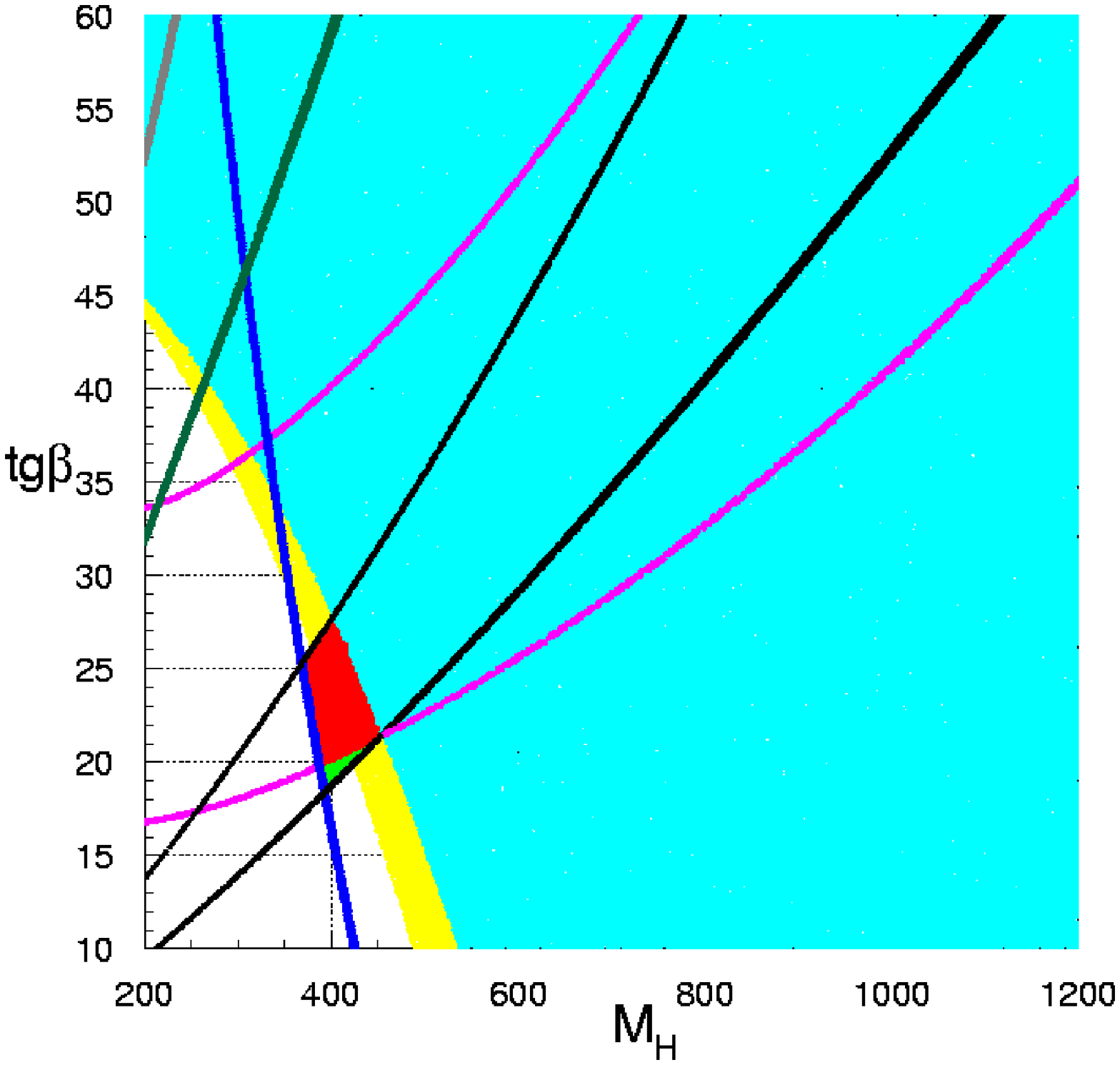}
\includegraphics[scale=0.38]{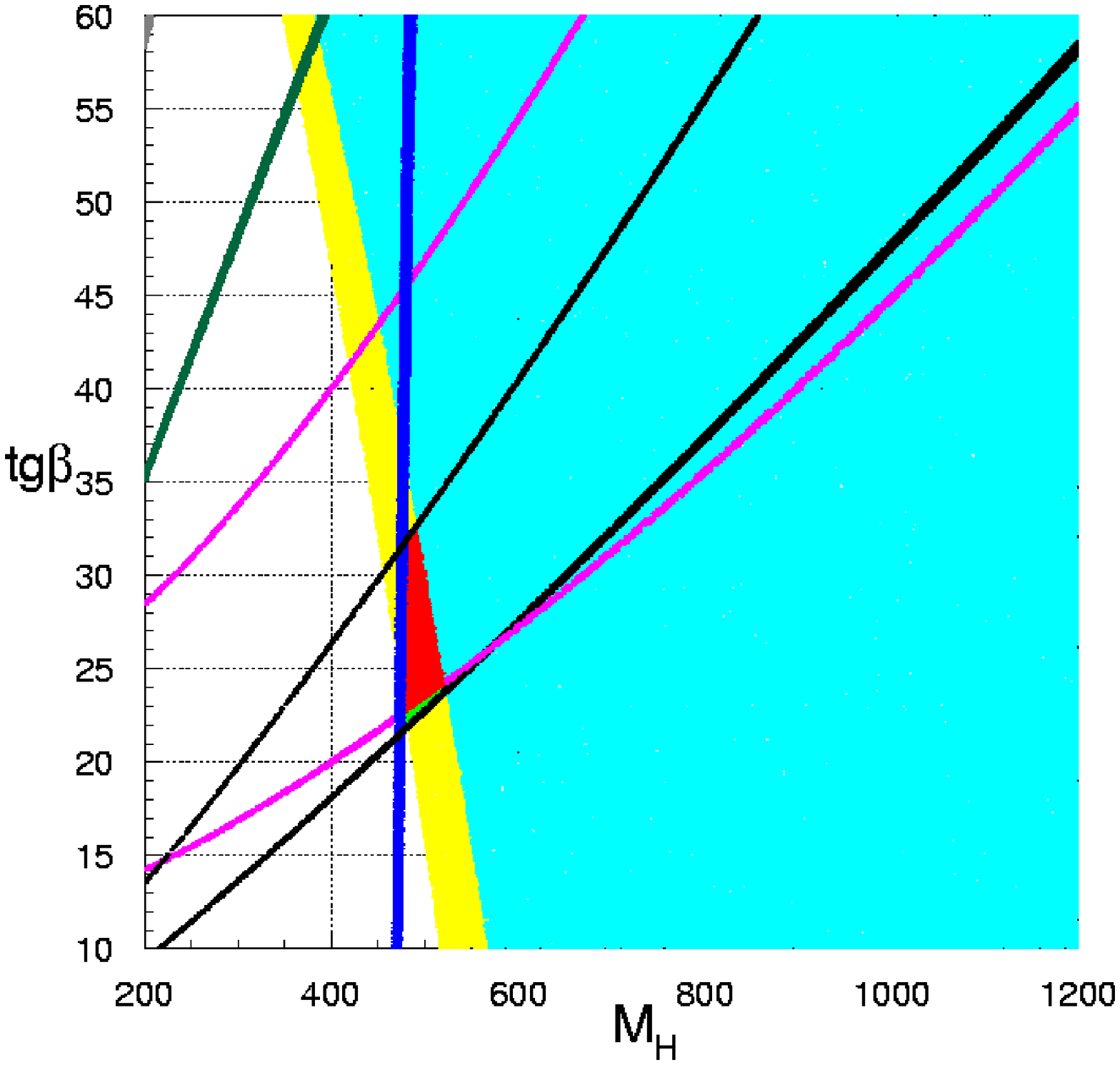}
\caption{\label{fig_MSSM_full_AP}
Same notations and conventions as in Figure~\ref{fig_MSSM_full}, but
for $A_U = 1$~TeV.}
\end{figure}

The combined constraints from low-energy observables and dark matter 
in the $\tan\beta$--$M_H$ plane
are illustrated in Figures~\ref{fig_MSSM_full} and~\ref{fig_MSSM_full_AP}.
The plots shown in these figures have been obtained setting  
$M_{\tilde q}=1.5$~TeV, $|A_U|=1$~TeV, $\mu=0.5$ or $1$~TeV, 
and $M_{\tilde \ell}=0.4$ or $0.3$~TeV.
The two sets of figures differ because of the sign of $A_U$.
The gaugino masses, satisfying the GUT condition $M_2\approx 2 M_1 \approx M_3/3$,
have been varied in each plot in order to fulfill the dark-matter conditions 
discussed in the previous Section (see Figure~\ref{fig:mu500}).
These conditions cannot be fulfilled in the gray (light-blue) areas with heavy $M_H$,
while the yellow band denotes the region where the 
stau coannihilation mechanism is active. 
The remaining bands correspond to the following constraints/reference-ranges 
from low-energy observables:\footnote{~For the sake of clarity, the resonance 
condition $M_H=2 M_1$ has been strictly enforced in the bands corresponding to 
the low-energy observables. Similarly, the stau coannihilation region has been 
determined imposing the relation $1< M_{\tilde{\tau}_R}/M_{\tilde{B}}<1.1$.}
\begin{itemize}
\item $B\to X_s \gamma$ [$1.01 < R_{ Bs\gamma} <1.24$]:  
allowed region between the two blue lines.
\item $a_\mu$ [$2 < 10^{9} (a_{\mu}^{\rm exp} - a_{\mu}^{\rm SM}) < 4$ \cite{gm2}]:
allowed region between the two purple lines.
\item{$ B\to \mu^+ \mu^-$}   [${\cal B}^{\rm exp}< 8.0 \times 10^{-8}$ \cite{Bmm}]: 
allowed region below the dark-green line.
\item{$\Delta M_{B_{s}}$}  [$\Delta M_{B_{s}} = 17.35 \pm 0.25~{\rm ps}^{-1}$ \cite{Dms}]:
allowed region below the gray line.
\item $B\to \tau \nu$ [$0.8 < R_{B\tau\nu} < 0.9$]: 
allowed region between the two black lines [ red (green) area if all the other 
conditions (but for $a_\mu$) are satisfied].
\end{itemize}

In the excluded regions at large $M_H$ (light-blue areas) 
the neutralino cannot satisfy the resonance condition 
$M_{\tchi_1} \approx M_H/2$ and, at the same time, be lighter than the sleptons. 
This is why the excluded regions become larger for lighter
$ M_{\tilde \ell}$. For the same reason, the 
excluded regions become larger for larger values of $\mu$
(we recall that $M^{2}_{\tilde{\tau}_{R}}\approx M^{2}_{\tilde \ell}-m_{\tau}\mu\tan\beta$).
We stress that in all cases we have explicitly checked the consistency with 
electroweak precision tests and the compatibility with exclusion bounds 
on direct SUSY searches. By construction, these conditions turn out to be 
naturally satisfied in the scenarios we have considered. The mot delicate 
constraint is the value of the lightest Higgs boson mass ($m_h$), which 
lies few GeV above its exlusion bound. In particular, we find  
118~GeV~$\leq m_h \leq$~120~GeV in the plots of Figure~\ref{fig_MSSM_full},
and  117~GeV~$\leq m_h \leq$~119~GeV in Figure~\ref{fig_MSSM_full_AP}.

\begin{figure}[t]
\begin{center}
\includegraphics[scale=0.50]{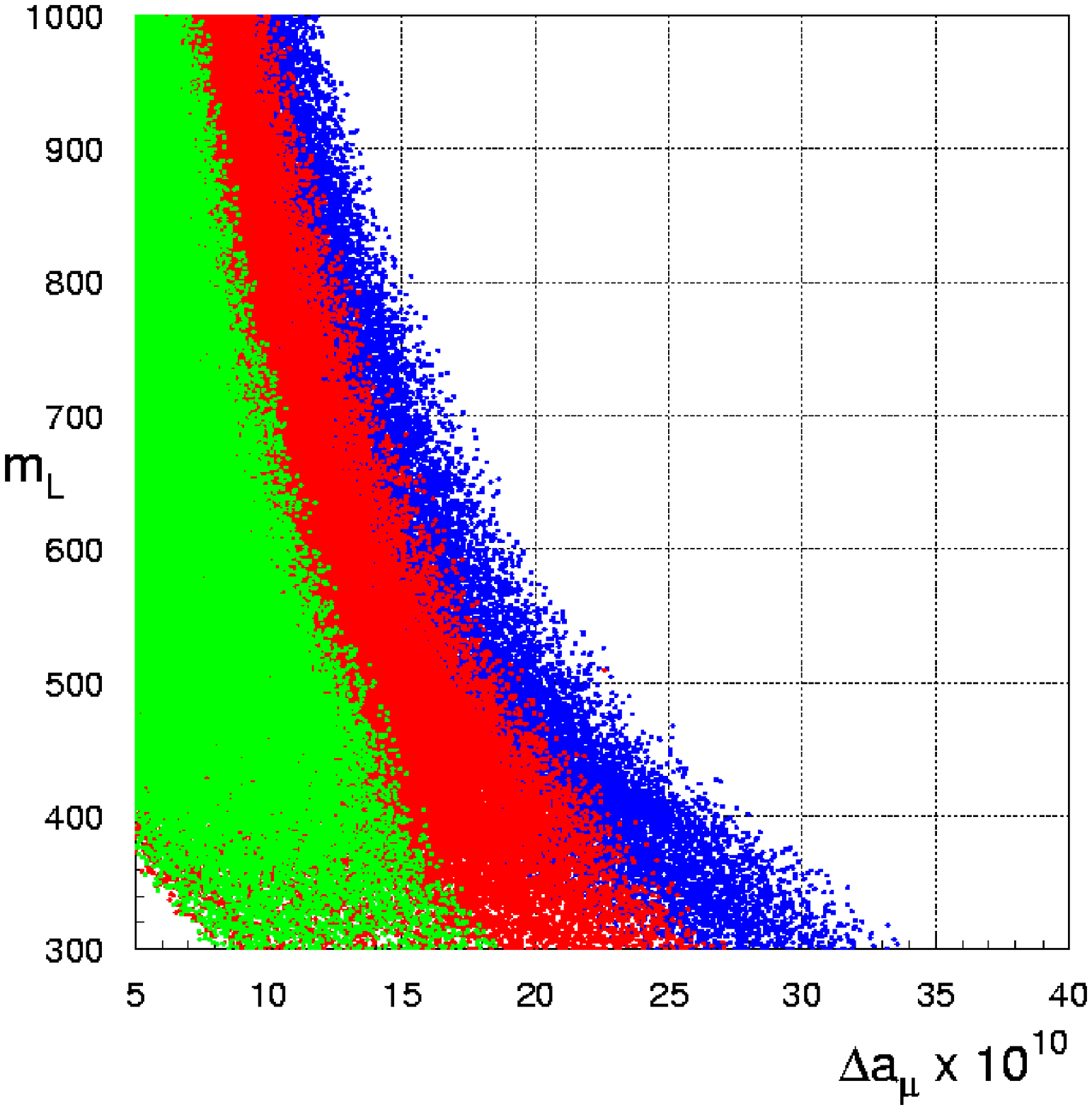}
\end{center}
\caption{\label{fig:gmvsbtn}
$\Delta a_\mu=(g_\mu-g^{\rm SM}_\mu)/2$ vs.~the slepton mass 
within the funnel region taking into account the $B\to X_s \gamma$ 
constraint and setting  
$R_{B\tau\nu}>0.7$  (blue), 
$R_{B\tau\nu}>0.8$  (red), 
$R_{B\tau\nu}>0.9$ (green).
The supersymmetric parameters 
have been varied in the following ranges: 200~GeV~$\leq M_{2}\leq$~1000~GeV,
500~GeV~$\leq \mu\leq$~1000~GeV, $10\leq \tan\beta \leq 50$. 
Moreover, we have set $A_U=-1$~TeV, $M_{\tilde{q}}=1.5~$TeV,
and imposed the GUT relation  $M_1 \approx M_2/2 \approx M_3/6$.
}
\end{figure}

As can be seen, in Figure~\ref{fig_MSSM_full}
the $B\to X_s \gamma$ constraint is always 
easily satisfied for $M_H \gsim 300$~GeV, or even 
lighter $M_H$ values for large $\tan\beta$ values.  
This is because the new range in Eq.~(\ref{eq:Bsgamma})
allows a significant (positive) non-standard contribution 
to the $B\to X_s \gamma$ rate. Moreover, having chosen $A_U<0$,
the positive charged-Higgs contribution is partially compensated
by the negative chargino-squarks amplitude. In 
Figure~\ref{fig_MSSM_full_AP}, where $A_U>0$, the $B\to X_s \gamma$
constraints are much more stringent and 
almost $\tan\beta$-independent. It is worth noting 
that in  Figure~\ref{fig_MSSM_full} the $B\to X_s \gamma$ 
information also exclude a region at large $M_H$: this is where the 
chargino-squarks amplitude dominates over the charged-Higgs
one, yielding a total negative corrections which is not 
favored by data. As already noted in \cite{IP}, 
the precise $\Delta M_{B_{s}}$ measurement and the
present limit on  $B\to \mu^+ \mu^-$ do not pose any 
significant constraint.

A part from the excluded region at large $M_H$, the 
most significant difference with respect to the analysis 
of Ref.~\cite{IP} (where dark-matter constraints 
have been ignored) is the interplay between 
$a_\mu$ and $B$-physics observables. The  correlation between 
$M_1$ and $M_H$ imposed by the dark matter constraint
is responsible for the rise with $M_H$ of the $a_\mu$ bands 
in Figures~\ref{fig_MSSM_full} and~\ref{fig_MSSM_full_AP}.
This makes more difficult to intercept the 
$B\to X_s \gamma$ and $\Btaun$ bands and, as a result, only a narrow 
area of the parameter space can fulfill all 
constraints. In particular, with the reference ranges
we have chosen, the best overlap occurs for moderate/large values of 
$\tan\beta$ and low values of $\mu$ and $M_{\tilde{\ell}}$.

On the other hand, we recall that the $\Btaun$
band in  Figure~\ref{fig_MSSM_full} 
does not correspond to the present experimental 
determination of this observable, but only to 
an exemplifying range. Assuming a stronger suppression 
of  $\BR(\Btaun)$ with respect to its SM value would 
allow a larger overlap between the 
$B\to X_s \gamma$ and $\Btaun$ bands in the regions with 
higher values of $\tan\beta$,  $\mu$ and $M_{\tilde{\ell}}$.
While if the  $\BR(\Btaun)$ measurement will converge toward the SM
value, for the reference values of $\mu$ and $M_{\tilde{\ell}}$
chosen in the figures ($\mu \geq 0.5$~GeV, $M_{\tilde{\ell}} \geq 0.3$~GeV) 
we deduce that: i)~for $R_{B\tau\nu}>0.8$  
the non-standard contribution to $a_{\mu}$ cannot not exceed $3 \times 10^{-9}$;
ii)~for $R_{B\tau\nu}>0.9$ the non-standard contribution to $a_{\mu}$ cannot not 
exceed $2 \times 10^{-9}$. An illustration of how the
non-standard contribution to $a_{\mu}$ varies as 
a funtion of $M_{\tilde{\ell}}$, imposing  different 
bounds on $R_{B\tau\nu}$, is shown in Figure~\ref{fig:gmvsbtn}.
Moreover, if the $\BR(\Btaun)$ measurement will converge toward the SM
value and the $a_\mu$ constraint is not considered, the green areas in 
Figures~\ref{fig_MSSM_full} and \ref{fig_MSSM_full_AP} are enlarged, allowing also
lower $\tan\beta$ values.

In short, the main result of this analysis is that in a scenario with heavy 
squarks and large trilinear couplings, the constraints and reference
ranges for the low-energy observables described above 
favor a charged Higgs mass in the $400-600$
GeV range and  $\tan\beta$ values in the 20-40 range. 
The structure of the favored $\tan\beta-M_H$ region depends on 
other SUSY parameters, mainly $\mu$ and $M_{\tilde \ell}$. Lower slepton masses shift the region toward  
lower $M_H$ and lower $\tan\beta$ values (in order 
to reproduce the $(g-2)_\mu$ anomaly and a neutralino LSP), while large
$\mu$ values reduce the favored region selecting larger $M_H$ and $\tan\beta$ values.

The analysis of future phenomenological signals of this scenario at LHC and other experiments
is beyond the scope of this work. However it should be noticed that both squarks and gluinos
are rather heavy (around or above $1$ TeV) and therefore not easily detectable.
On the other hand, a direct detection of the charged Higgs and/or of the sleptons 
should be possible. In this case, the combination of high-energy and low-energy observables 
would allow to determine the $\tan\beta$ parameter very precisely.

\subsection{Correlation between LFV decays and $(g-2)_{\mu}$}

As we have seen from the analysis of Figures~\ref{fig_MSSM_full} 
and~\ref{fig_MSSM_full_AP}, a key element which characterizes the 
scenario we are considering is the interplay between  $(g-2)_{\mu}$ and 
$B$-physics observables. Since $(g-2)_{\mu}$ is affected by
irreducible theoretical uncertainties \cite{gm2}, it is desirable
to identify additional observables sensitive to the same (or a very similar)
combination of supersymmetric parameters. An interesting possibility is 
provided by the LFV transitions 
$\ell_i \rightarrow \ell_j\gamma$ and, in particular, by the 
$\mu \rightarrow e \gamma$ decay.
Apart from the unknown overall normalization associated to the LFV couplings,
the amplitude of these transitions are closely connected
to those generating the non-standard contribution to $a_{\mu}$ \cite{Hisa1}.

LFV couplings naturally appear in the MSSM once we extend 
it to accommodate the non-vanishing neutrino masses 
and mixing angles by means of a supersymmetric seesaw mechanism~\cite{fbam}. 
In particular, the renormalization-group-induced LFV entries 
appearing in the left-handed slepton mass matrices have the following 
form \cite{fbam}:
\be
\delta_{LL}^{ij} ~=~ 
\frac{ \left( M^2_{\tilde \ell} \right)_{L_i L_j}}
{\sqrt{\left( M^2_{\tilde \ell} \right)_{L_i L_i} 
\left( M^2_{\tilde \ell} \right)_{L_j L_j}}} =
c_\nu (Y^\dagger_\nu Y_\nu)_{ij}~,
\label{eq:fbam}
\ee
where  $Y_\nu$ are the neutrino Yukawa couplings and 
$c_\nu$ is a numerical coefficient, depending 
on the SUSY spectrum, typically of  $\cO(0.1$--$1)$.
As is well known, the information from neutrino 
masses is not sufficient to determine in a model-independent 
way all the seesaw parameters relevant to LFV rates and,
in particular, the neutrino Yukawa couplings. 
To reduce the number of free parameters specific SUSY-GUT 
models and/or flavour symmetries need to be employed.
Two main roads are often considered in the literature 
(see e.g.~Ref.~\cite{Masierorew} and references there in):
the case where the charged-lepton LFV couplings are linked 
to the CKM matrix (the quark mixing matrix) and the case where 
they are connected to the PMNS matrix (the neutrino mixing matrix). 
These two possibilities can be formulated in terms of
well-defined flavour-symmetry structures starting from the 
MFV hypothesis \cite{MLFV,MFVGUT}. A useful reference scenario
is provided by the so-called MLFV hypothesis~\cite{MLFV},
namely by the assumption that the flavour degeneracy in the lepton sector 
is broken only by the neutrino Yukawa couplings, in close analogy to the quark sector.
According to this hypothesis, the LFV entries introduced in Eq.~(\ref{eq:fbam})
assume the following form 
\be
\delta_{LL}^{ij} ~=~ 
c_\nu (Y^\dagger_\nu Y_\nu)_{ij}  ~\approx~   c_{\nu}\dis\frac{m_\nu^{\rm atm} M_{\nu_R} 
}{v_2^2}~ U_{i 3} U^*_{j 3} 
\label{eq:delta_LL}
\ee
where $M_{\nu_R}$ is the average right-handed neutrino mass and
$U$ denote the PMNS matrix.

Once non-vanishing LFV entries in the slepton mass matrices 
are generated, LFV rare decays are naturally induced by
one-loop diagrams with the exchange of gauginos and sleptons 
(gauge-mediated LFV amplitudes).
\footnote{~An additional and potentially large class of LFV 
contributions to rare decays comes from the Higgs sector
through the effective LFV Yukawa interactions induced by 
non-holomorphic terms \cite{bkl}.
However, these effects become competitive with the
gauge-mediated ones only if $\tan\beta \sim \mathcal{O}(40-50)$ and
if the Higgs masses are roughly one order of magnitude lighter then 
the slepton masses \cite{paradisiH}.
Since we consider a slepton mass spectrum well below the TeV scale,
Higgs mediated LFV effects do not play a relevant role in our 
analysis.} In particular, the leading contribution due to the exchange 
of charginos, leads to
\beq
\frac{\BR(\ell_i\rightarrow \ell_j\gamma)}
{\BR(\ell_{i}\rightarrow \ell_{j}\nu_{\ell_i}\bar{\nu_{\ell_j}})} =
\frac{48\pi^{3}\alpha}{G_{F}^{2}} \left| 
\frac{\alpha_{2}}{4\pi}
\left(\frac{\mu M_{2}}{m^2_{L}}\right)
\frac{f_{2c}\left( M^{2}_{2}/M^{2}_{\tilde \ell}, \mu^2/M^{2}_{\tilde \ell} \right) }{
(M_{2}^2\!-\!\mu^2)}\,\delta_{LL}^{ij} \tan\beta,
\right|^2
\label{eq:llg}
\eeq
where the loop function $f_{2c}(x,y)$ is defined as $f_{2c}(x,y)=f_{2c}(x)-f_{2c}(y)$
in terms of 
\beq
f_{2c}(a)=\frac{-a^2-4a+5+2(2a+1)\ln a}{2(1-a)^4}\,.
\eeq

\begin{figure}[t]
\centering
\includegraphics[scale=0.38]{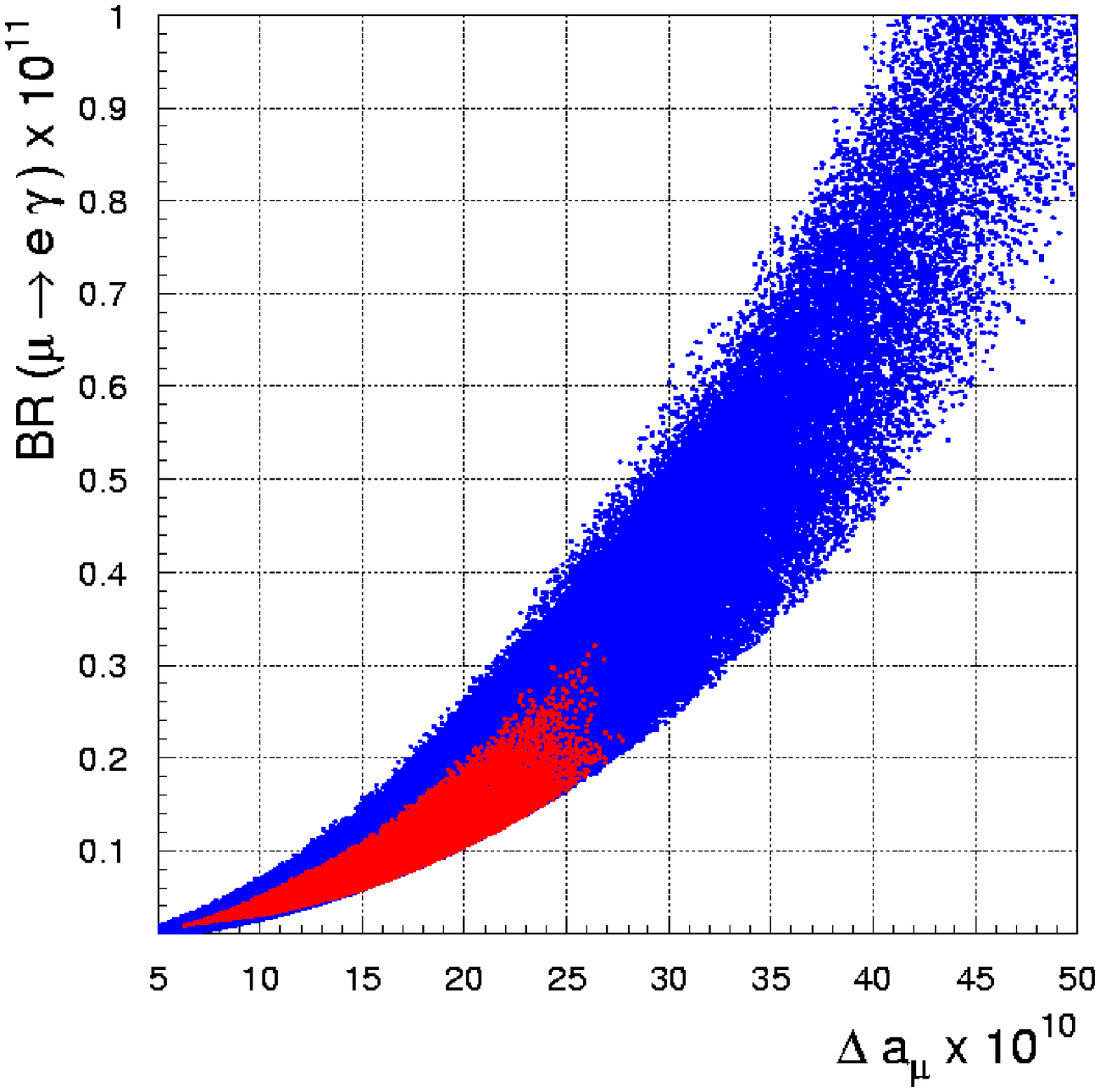}
\includegraphics[scale=0.38]{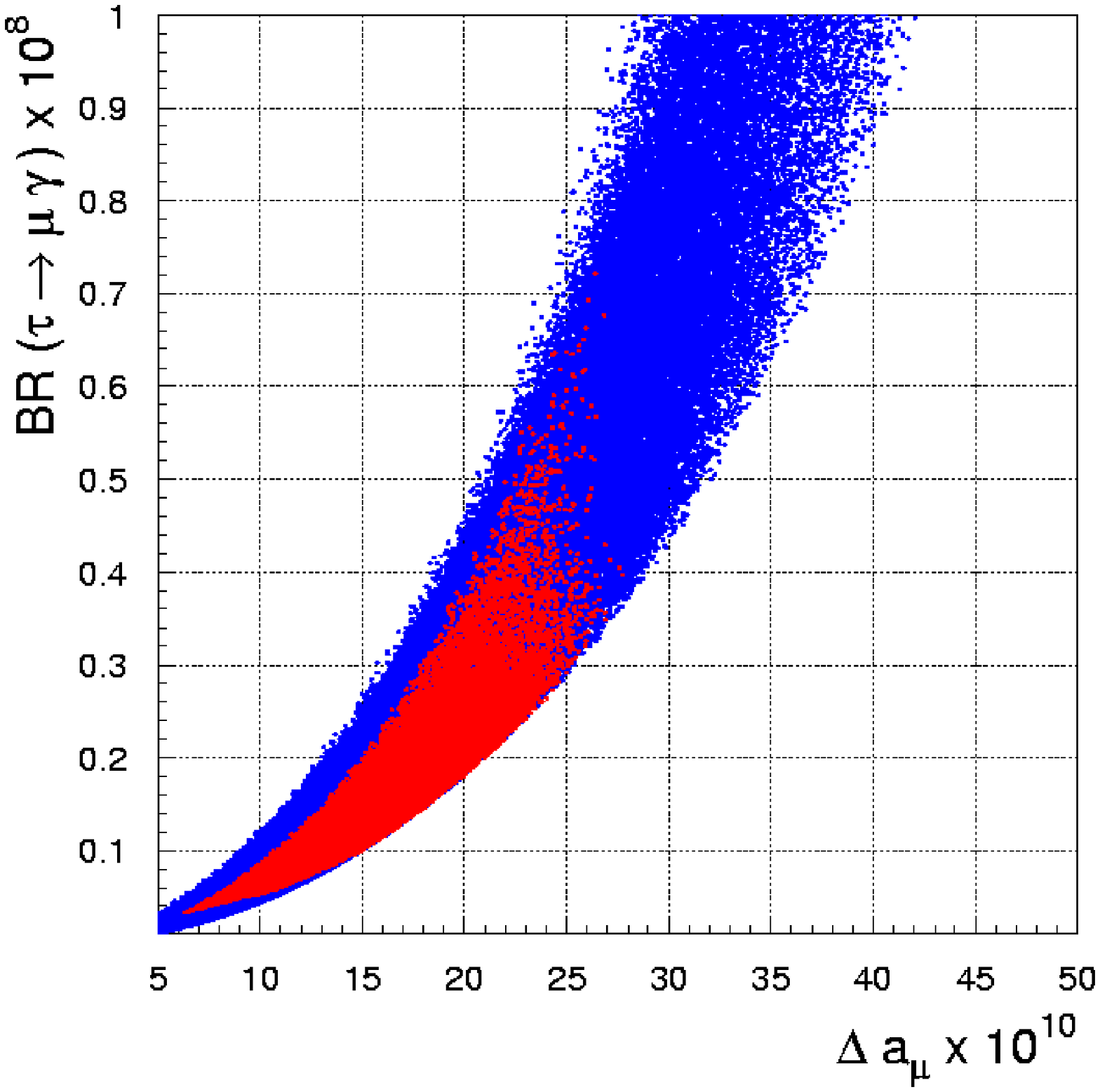}
\caption{\label{fig2}
Expectations for  $\cB(\mu\rightarrow e\gamma)$ and 
$\cB(\tau\rightarrow \mu\gamma)$ vs.~$\Delta a_\mu=(g_\mu-g^{\rm SM}_\mu)/2$, 
assuming $|\delta_{LL}^{12}|=10^{-4}$ and  $|\delta_{LL}^{23}|=10^{-2}$.
The plots have been obtained employing the following ranges:
300~GeV~$\leq M_{\tilde{\ell}}\leq$~600~GeV, 200~GeV~$\leq M_{2}\leq$~1000~GeV,
500~GeV~$\leq \mu\leq$~1000~GeV, $10\leq \tan\beta \leq 50$, and 
setting $A_U=-1$~TeV, $M_{\tilde{q}}=1.5~$TeV.
Moreover, the GUT relations
$M_2\approx 2M_1$ and  $M_3\approx 6M_1$ are assumed.
The red areas correspond to points within the funnel region which 
satisfy the $B$-physics constraints listed in Section~\ref{sect:combined}
[$\cB(B_s\rightarrow \mu^+\mu^-)<8\times 10^{-8}$,
$1.01 <R_{B s \gamma}<1.24$, $0.8<R_{B\tau\nu}<0.9$, 
$\Delta M_{B_{s}} = 17.35 \pm 0.25~{\rm ps}^{-1}$].}
\end{figure}

\begin{figure}[p]
\includegraphics[scale=0.38]{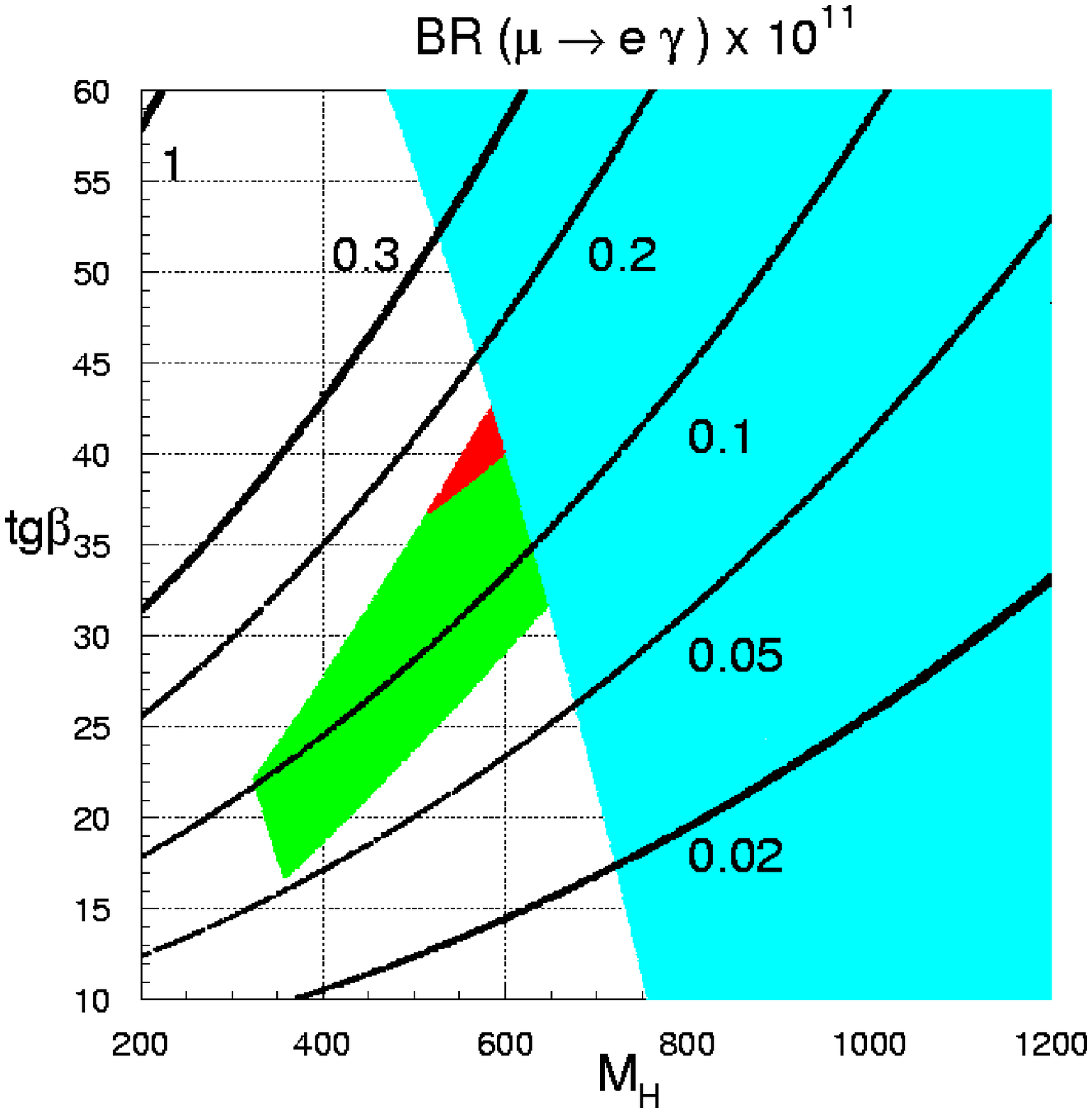}
\includegraphics[scale=0.38]{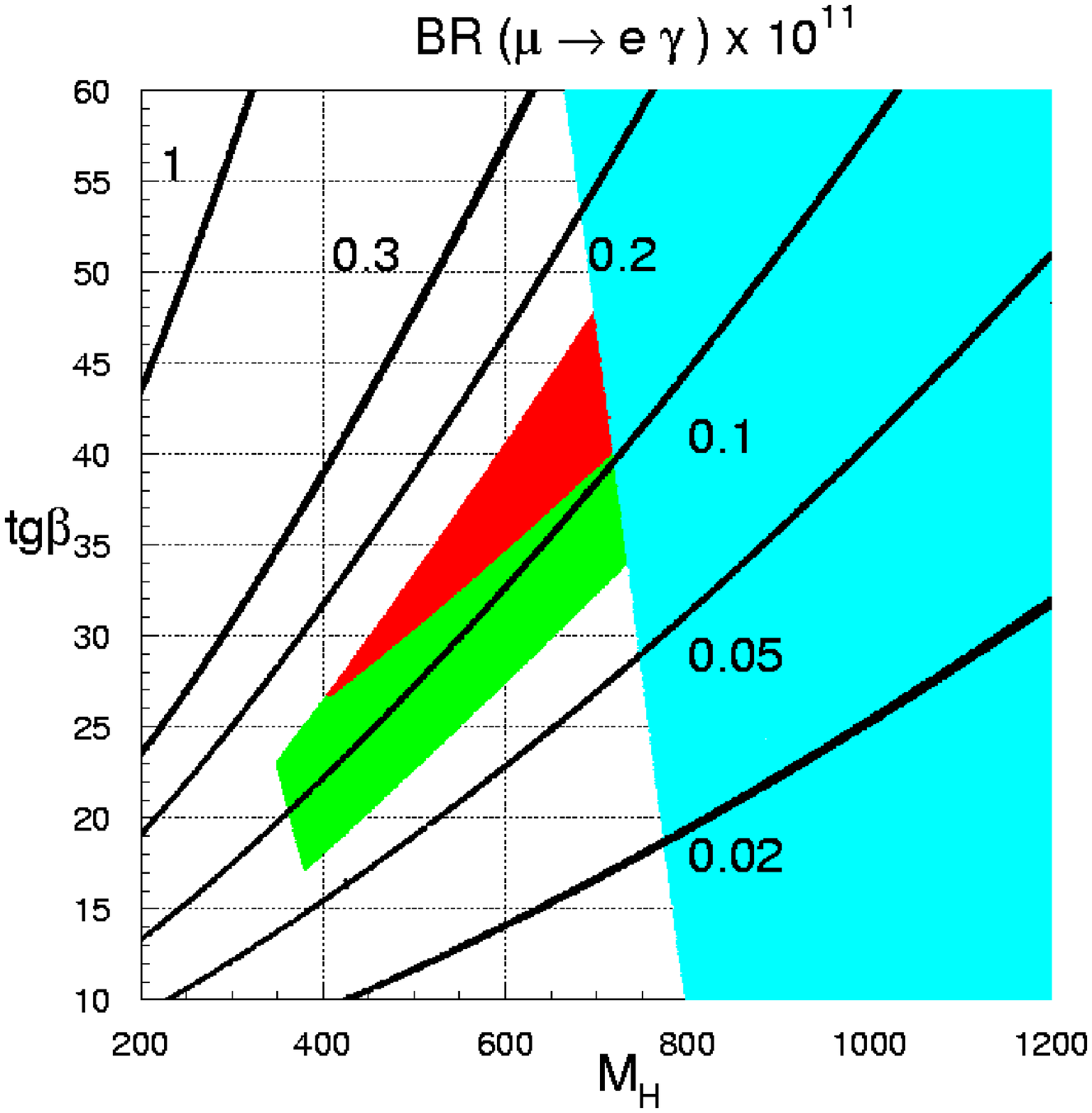} \\
\includegraphics[scale=0.38]{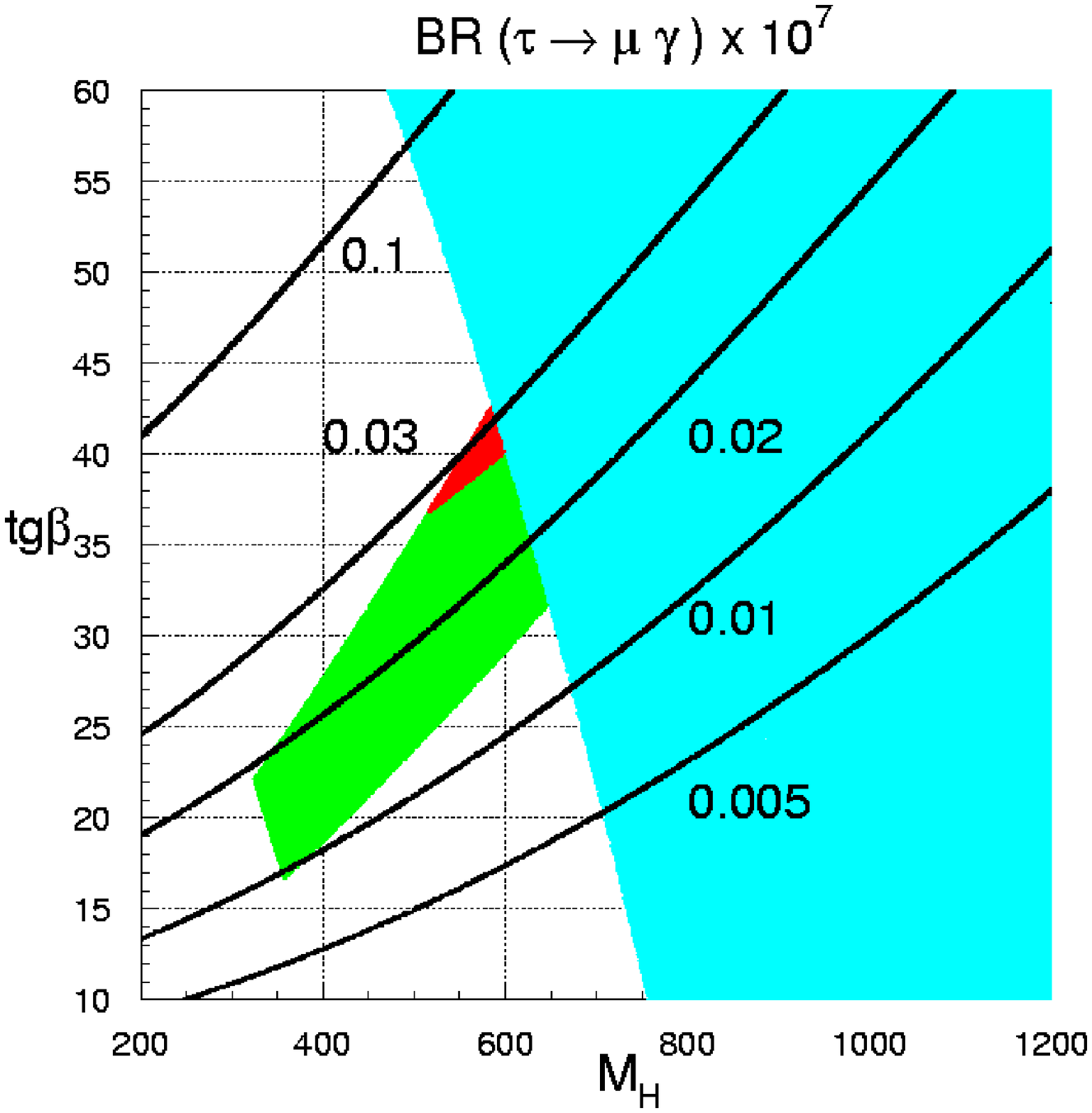}
\includegraphics[scale=0.38]{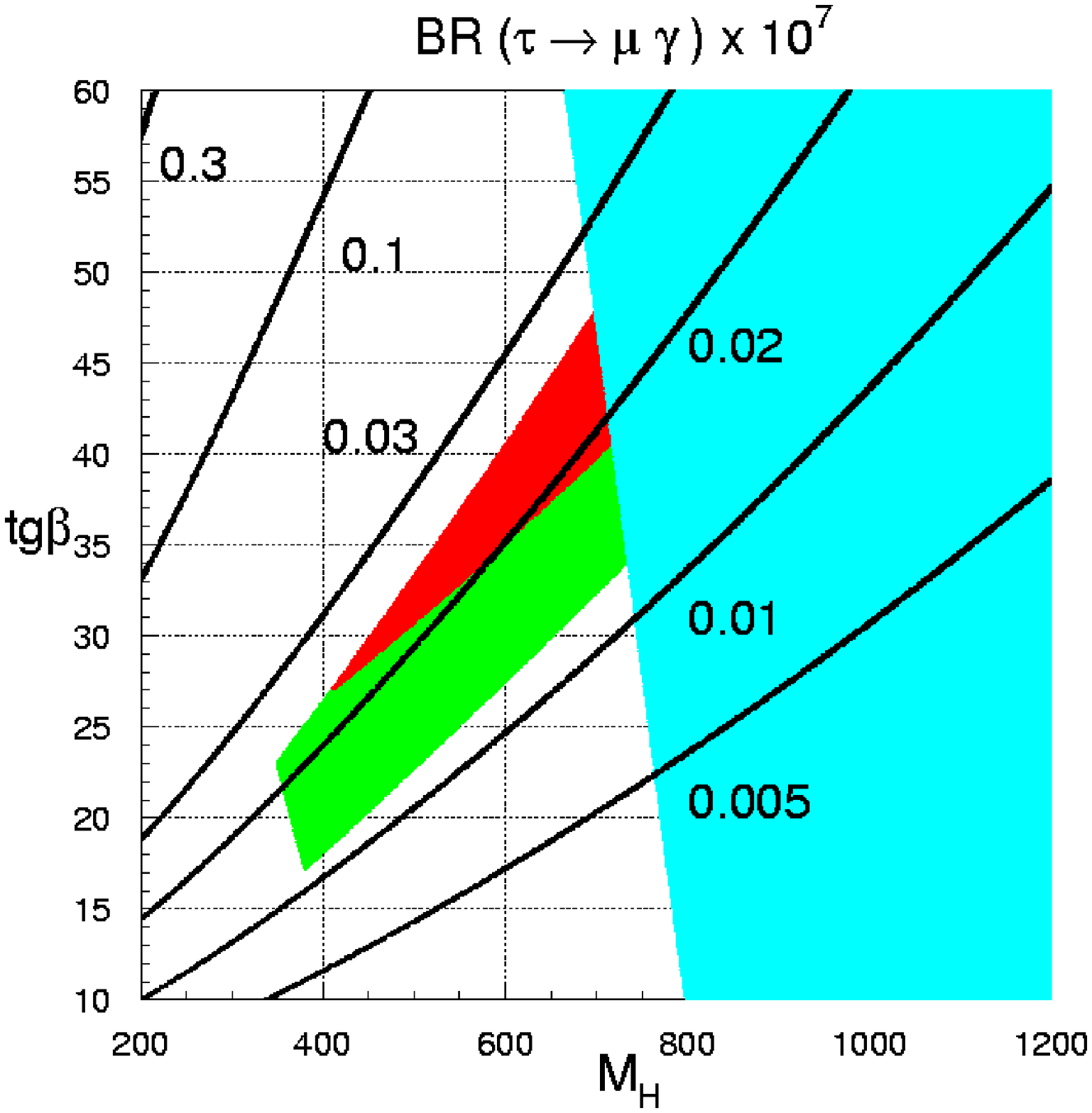}
\caption{\label{fig_meg_full} Isolevel curves for 
$\cB(\mu\rightarrow e\gamma)$ and 
$\cB(\tau\rightarrow \mu\gamma)$
assuming $|\delta_{LL}^{12}|=10^{-4}$ and  $|\delta_{LL}^{23}|=10^{-2}$
in the  $\tan\beta$--$M_H$ plane. The green/red areas correspond
to the allowed regions for the low-energy observables 
illustrated in Figure~\ref{fig_MSSM_full} for 
$[\mu,M_{\tilde{\ell}}]=[1.0,0.4]$~TeV (left plots), 
$[\mu,M_{\tilde{\ell}}]=[0.5,0.4]$~TeV (right plots).}
\end{figure}

Given that both $\ell_{i}\rightarrow\ell_{j}\gamma$ and $\Delta a_\mu=(g_\mu-g^{\rm SM}_\mu)/2$
are generated by dipole operators, it is natural to establish a link between them.
To this purpose, we recall the dominant contribution to $\Delta a_\mu$ is also provided 
by the chargino exchange and can be written as
\beq
\Delta a_{\mu} = -\frac{\alpha_{2}}{4 \pi}m_{\mu}^{2}
\left(\frac{\mu M_2}{m_{L}^{2}}\right)
\frac{g_{2c}\left( M^{2}_{2}/M^{2}_{\tilde \ell}, \mu^2/M^{2}_{\tilde \ell} \right) }{
(M_{2}^2\!-\!\mu^2)}
\tan\beta\,,
\label{eq:dmu}
\eeq
with $g_{c2}(x,y)$ defined as $f_{c2}(x,y)$ in terms of 
\beq
g_{c2}(a) = \frac{(3-4a+a^2+2\log a )}{(a-1)^3}~.
\eeq
It is then straightforward to deduce the relation 
\beqa
\frac{\BR(\ell_i\rightarrow \ell_j\gamma)}
{\BR(\ell_i\rightarrow \ell_j\nu_{\ell_i}\bar{\nu_{\ell_j}})} =
\frac{48\pi^{3}\alpha}{G_{F}^{2}}
\left[\frac{\Delta a_{\mu}}{m_{\mu}^{2}}\right]^{2}\,
\left[\frac{f_{2c}\left( M^{2}_{2}/M^{2}_{\tilde \ell}, \mu^2/M^{2}_{\tilde \ell} \right)}{
g_{2c}\left( M^{2}_{2}/M^{2}_{\tilde \ell}, \mu^2/M^{2}_{\tilde \ell} \right)}
\right]^2
\,\left| \delta_{LL}^{ij} \right|^2~.
\label{eq:ratio_LFV}
\eeqa
To understand the relative size of the correlation, 
in the limit of degenerate SUSY spectrum we get
\beqa
\BR(\ell_i\rightarrow \ell_j \gamma) ~\approx~ 
\left[\frac{\Delta a_{\mu}}{ 20 \times 10^{-10}}\right]^{2} \times \left\{
\ba{ll}
1 \times 10^{-4} \, \left| \delta_{LL}^{12} \right|^2 \qquad  & [\mu\to e]~,  \\
2 \times 10^{-5} \, \left| \delta_{LL}^{23} \right|^2         & [\tau\to \mu]~.  
\ea
\right.
\eeqa
A more detailed analysis of the stringent correlation
between the $\ell_i \rightarrow \ell_j \gamma$ transitions  
and $\Delta a_\mu$ in our scenario is illustrated in Fig.\ref{fig2}. 
Since the loop functions for the two processes are not
identical, the correlation is not exactly a line; however, 
it is clear that the two observables are closely connected. 
We stress that the numerical results shown in Fig.\ref{fig2}.
have been obtained using the exact formulae reported in Ref.~\cite{hisano} 
for the supersymmetric contributions to 
both $\BR(\ell_i \rightarrow \ell_j \gamma)$  and $\Delta a_\mu$
(the simplified results in the mass-insertion approximations 
in Eqs.~(\ref{eq:llg})--(\ref{eq:ratio_LFV})
have been shown only for the sake of clarity).
The red areas are the regions where the $B$-physics constraints
are fulfilled. In our scenario the $B$-physics constraints put a
lower bound on $M_H$ and therefore, through the funnel-region relation,
also on $M_{1,2}$ (see Figs.~\ref{fig_MSSM_full} and~\ref{fig_MSSM_full_AP}).
As a result, the allowed
ranges for $\Delta a_\mu$ and $\BR(\ell_i \rightarrow \ell_j \gamma)$ are 
correspondingly lowered. A complementary illustration of the 
interplay of $B$ physics observables, dark-matter constraints, 
 $\Delta a_\mu$, and LFV rates --within our scenario-- 
is shown in Figure~\ref{fig_meg_full}.\footnote{~For comparison, 
a detailed study of LFV transitions imposing dark-matter constraints
--within the constrained MSSM with right-handed neutrinos--
can be found in Ref.~\cite{lfvwmap}.}

\begin{table}[t]
\begin{center}
\begin{tabular}{||l|l||c|c||}
\hline
Observable &  Exp. bound &   Bound  on   & Expected $|\delta_{LL}|$ in MLFV  \\
           &             &  the eff coupl.   & for~$M_{\nu_R}=10^{12}$~GeV \\
\hline
  $\BR(\mu \to e\,\gamma)$    &  $ < 1.2\times 10^{-11}$  
  & $|\delta_{LL}^{21}|< 3\times 10^{-4}$  & $(0.3-3)\times10^{-4}$     \\
  $\BR(\tau \to e\,\gamma)$   &  $ < 1.1\times 10^{-7}$   
  & $|\delta_{LL}^{31}|< 8\times 10^{-2}$  & $(0.3-3)\times10^{-4}$ \\
  $\BR(\tau \to \mu\,\gamma)$ &  $ < 6.8\times 10^{-8}$   
  & $|\delta_{LL}^{32}|< 6\times 10^{-2}$  & $0.8\times10^{-3}$  \\
\hline 
\end{tabular}
\end{center}
\caption{\label{lfvtable}
Present experimental bounds on the radiative LFV decays 
of $\tau$ and $\mu$ leptons \cite{Yao:2006px} and corresponding bounds 
on the effective LFV couplings $\delta^{ij}_{LL}$. 
The bounds are obtained by means of  Eq.~(\ref{eq:ratio_LFV})
setting $\Delta a_{\mu} = 20 \times 10^{-10}$. 
The expectations for the $\delta^{ij}_{LL}$ reported in the last two 
columns correspond to MLFV ansatz in  Eq.~(\ref{eq:delta_LL}) with 
$c_\nu=1$ and  $M_{\nu_R}=10^{12}$~GeV.  }
\end{table}

The normalization $|\delta_{LL}^{12}|=10^{-4}$ used in 
Figures~\ref{fig2} and~\ref{fig_meg_full}
corresponds to the central value in Eq.~(\ref{eq:delta_LL}) 
for $c_\nu=1$ and  $M_{\nu_R}=10^{12}$~GeV. 
This normalization can be regarded as a rather natural (or even pessimistic) choice.
\footnote{~For $M_{\nu_R}\ll 10^{12}$~GeV other sources of LFV, such as the quark-induced 
terms in Grand Unified Theories cannot be neglected \cite{BH}.
As a result, in many realistic scenarios it 
is not easy to suppress LFV entries in the slepton mass matrices 
below the $10^{-4}$ level~\cite{MFVGUT}.} 
As can be seen from  Figures~\ref{fig2} and~\ref{fig_meg_full},
for such natural choice of $\delta_{LL}$ 
the $\mu\rightarrow e\gamma$ branching ratio is in the $10^{-12}$ 
range, i.e.~well within the reach of MEG~\cite{MEG} experiment.
Note that this is a well-defined prediction of our scenario, 
where the connection between  $\mu\rightarrow e\gamma$ 
and $\Delta a_\mu$ allows us to substantially reduce the 
number of free parameters. In particular, the requirement of  
a supersymmetric contribution to $\Delta a_\mu$ of $O(10^{-9})$
forces a relatively light sparticle spectrum 
and moderate/large $\tan\beta$ values which both tend to 
enhance the LFV rates. This fact already 
allows to exclude values of $\delta_{LL}^{12}$ above  $10^{-3}$, 
for which $\BR(\mu\rightarrow e\gamma)$ would exceed 
the present experimental bound.\footnote{~For a recent and detailed analysis on the bounds for LFV 
soft breaking term as functions of the relevant SUSY parameters
(without assuming the present $g-2$ anomaly as a hint of New Physics),
see Ref.\cite{Paride}.}
Within the MLFV hypothesis, this translates into a non-trivial upper bound 
on the right-handed neutrino mass: $M_{\nu_R} < 10^{13}$~GeV.

On the other hand, the normalization $|\delta_{LL}^{23}|=10^{-2}$
adopted for the $\tau\to\mu\gamma$ mode is more optimistic 
given the MLFV expectations in  Table~\ref{lfvtable}. 
We have chosen this reference value because only for 
such large LFV entries the $\tau\to\mu\gamma$ transition 
could be observed in the near future. From the comparison of 
Figure~\ref{fig2} and  Table~\ref{lfvtable} we deduce that, 
unless $\mu\to e\gamma$ is just below its present exclusion bound, 
an observation of $\tau\to\mu\gamma$ above $10^{-9}$ would exclude the LFV 
pattern predicted by the MLFV hypothesis~\cite{MLFV}.


\section{Conclusions}

Within the wide parameter space of the  supersymmetric 
extensions of the SM, the regime of large $\tan\beta$ and heavy squarks 
represents an interesting corner. It is a region consistent with present data, 
where the $(g-2)_\mu$ anomaly and the upper bound on the Higgs boson mass
could find a natural explanation. Moreover, this region could possibly be 
excluded or gain more credit with more precise data on a few $B$-physics 
observables, such as $\BR(\Btaun)$ and   $\BR(B \to \ell^+\ell^-)$.
In this paper we have analysed the correlations 
of the most interesting low-energy observables within this scenario, 
interpreting the $(g-2)_\mu$ anomaly 
as the first hint of this scenario, and assuming that 
the relic density of a Bino-like LSP 
accommodates the observed dark matter distribution. In view of 
improved experimental searches of LFV decays, we have also analysed
the expectations for the rare decays  $\mu\to e\gamma$ 
and $\tau\to\mu(e) \gamma$  in this framework.

The main conclusions of our analysis can be summarised as follows:
\begin{itemize}
\item 
Within this region it is quite natural to fulfill the dark-matter
constraints thanks to the resonance enhancement of the 
$\tchi_1 \tchi_1 \to H,A \to f\bar f$ cross section ($A$-funnel region). 
As shown in Fig.~\ref{fig:mu500}, this mechanism is successful in a
sufficiently wide area of the parameter space.
\item  
From the phenomenological point of view, 
the most significant impact of the dark-matter constraints 
is the non-trivial interplay between $a_\mu$ and the 
 $B$-physics observables. 
A supersymmetric contribution to $a_\mu$
of $\cO(10^{-9})$
is perfectly compatible with the present constraints 
from $\BR(B \to X_s \gamma)$, especially for $A_U <0$. 
However, taking into account the correlation between 
neutralino and charged-Higgs masses occurring in the 
$A$-funnel region, this implies a sizable 
suppression of  $\BR(\Btaun)$ with respect to its SM prediction. 
As shown in Figure~\ref{fig:gmvsbtn},
the size of this suppression depends on the 
slepton mass, which in turn controll the size of the 
supersymmetric contribution to $a_\mu$. 
In particular, 
we find that $\Delta a_{\mu} \gsim 2 \times 10^{-9}$ implies 
a relative suppression of $\BR(\Btaun)$ larger than $10\%$
A more precise determination of $\BR(\Btaun)$ is therefore a 
key element to test this scenario. 
\item  
A general feature of supersymmetric models is a strong 
correlation between $\Delta a_{\mu}$ and the rate of the
LFV transitions $\ell_i \rightarrow \ell_j\gamma$ \cite{Hisa1}. 
We have re-analysed this correlation in our framework, taking into account 
the updated constraints on  $\Delta a_{\mu}$ and  $B$-physics 
observables, and employing the MLFV ansatz \cite{MLFV} 
to relate the flavour-violating entries in the slepton 
mass matrices to the observed neutrino mass matrix. 
According to the latter (pessimistic) hypothesis, we find 
that the $\mu\rightarrow e\gamma$ branching ratio is likely to 
be within the reach of MEG~\cite{MEG} experiment,
while LFV decays of the $\tau$ leptons are unlikely to 
exceed the $10^{-9}$ level.
\end{itemize}

\section*{Acknowledgments} 
We thank Uli Haisch, Enrico Lunghi, and Oscar Vives  
for useful discussions. This work is supported in part 
by the EU Contract No.~MRTN-CT-2006-035482, ``FLAVIAnet''.
P.P.~acknowledges the support of the spanish MEC and FEDER 
under grant FPA2005-01678.

\end{document}